\documentclass{article}

\usepackage[english]{babel}

\usepackage[a4paper,top=2cm,bottom=2cm,left=3cm,right=3cm,marginparwidth=1.75cm]{geometry}

\usepackage{amsmath}
\usepackage{graphicx,comment}
\usepackage[table]{xcolor}
\usepackage{url} 
\usepackage{dsfont}
\usepackage{cite}
\usepackage{tabularx,multirow}

\usepackage{amsmath,amssymb,amsfonts}
\usepackage{algorithmic}
\usepackage{subfig}
\usepackage{textcomp}
\usepackage{multirow}
\usepackage{etoolbox}

\usepackage[binary-units]{siunitx}

\graphicspath{{Figures/}}

\usepackage[acronym]{glossaries}
\newacronym{stp}{STP}{short-term plasticity}
\newacronym{stf}{STF}{short-term facilitation}
\newacronym{std}{STD}{short term depression}
\newacronym{rram}{RRAM}{resistive random access memory}
\newacronym{cpu}{CPU}{central processing unit}
\newacronym{cmos}{CMOS}{complementary metal oxide semiconductor}
\newacronym{sram}{SRAM}{static random access memory}
\newacronym{epsc}{EPSC}{excitatory post-synaptic current}
\newacronym{if}{IF}{integrate-and-fire}
\newacronym{lif}{LIF}{leaky integrate-and-fire}
\newacronym{wm}{WM}{working memory}
\newacronym{rnn}{RNN}{recurrent neural network}
\newacronym{lstm}{LSTM}{long-short term memory}
\newacronym{gru}{GRU}{gated recurrent unit}
\newacronym{gpu}{GPU}{graphic processing unit}
\newacronym{fpga}{FPGA}{field programmable gate array}
\newacronym{asic}{ASIC}{application specific integrated circuit}
\newacronym{cf}{CF}{conductive filament}
\newacronym{ml}{ML}{machine learning}
\newacronym{lrs}{LRS}{low resistive state}
\newacronym{hrs}{HRS}{high resistive state}
\newacronym{te}{TE}{top electrode}
\newacronym{be}{BE}{bottom electrode}
\newacronym{snr}{SNR}{signal-to-noise ratio}

\title{Tunable Synaptic Working Memory with Volatile Memristive Devices}

\author{Saverio Ricci*$^{1}$, David Kappel*$^{2,3}$, Christian Tetzlaff$^{2,4}$,\\ Daniele Ielmini$^{1}$, and Erika Covi$^{5}$\\[4mm]
\begin{minipage}[t]{0.48\linewidth}
\centering
\small $^{1}$\;Dipartimento di Elettronica, Informazione e Bioingegneria (DEIB)\\
\small Politecnico di Milano\\
\small Piazza Leonardo da Vinci 32, 20133 Milano, Italy
\end{minipage}
\begin{minipage}[t]{0.48\linewidth}
\centering
\small $^{2}$\;Bernstein Center for Computational Neuroscience\\
\small III Physikalisches Institut – Biophysik\\
\small Georg-August Universit\"at\\
\small Friedrich-Hund-Platz 1, 37077 G\"ottingen, Germany
\end{minipage}\vspace{6mm}\\
\begin{minipage}[t]{0.48\linewidth}
\centering
\small $^{3}$\;Institut für Neuroinformatik \\
\small Ruhr-Universit\"at Bochum \\
\small Universit\"atsstr. 150 NB 3/32\\
44801 Bochum, Germany
\end{minipage}
\begin{minipage}[t]{0.48\linewidth}
\centering
\small $^{4}$\;Group of Computational Synaptic Physiology\\
\small Department for Neuro- and Sensory Physiology\\
\small University Medical Center G\"ottingen\\
\small Von-Siebold-Str. 3A, 37075 G\"ottingen, Germany
\end{minipage}\vspace{6mm}\\
\small $^{5}$\;NaMLab gGmbH\\
\small N\"othnitzer Str. 64 a \\
\small 01187 Dresden, Germany
\vspace{6mm}\\
\small* These Authors contributed equally to this work.}

\date{}

\begin{document}
\maketitle

\begin{abstract} 
Different real-world cognitive tasks evolve on different relevant timescales. Processing these tasks requires memory mechanisms able to match their specific time constants. In particular, the working memory utilizes mechanisms that span orders of magnitudes of timescales, from milliseconds to seconds or even minutes. This plentitude of timescales is an essential ingredient of working memory tasks like visual or language processing.  
This degree of flexibility is challenging in analog computing hardware because it requires the integration of several reconfigurable capacitors of different size. Emerging volatile memristive devices present a compact and appealing solution to reproduce reconfigurable temporal dynamics in a neuromorphic network.

We present a demonstration of working memory using a silver-based memristive device whose key parameters, retention time and switching probability, can be electrically tuned and adapted to the task at hand. First, we demonstrate the principles of working memory in a small scale hardware to execute an associative memory task. Then, we use the experimental data in two larger scale simulations, the first featuring working memory in a biological environment, the second demonstrating associative symbolic working memory.

\end{abstract}

\newpage
\begingroup
\let\clearpage\relax
\section{Introduction}
\label{sec:intro}

\begin{figure}[ht]
    \centering
    \includegraphics[width=0.95\textwidth]{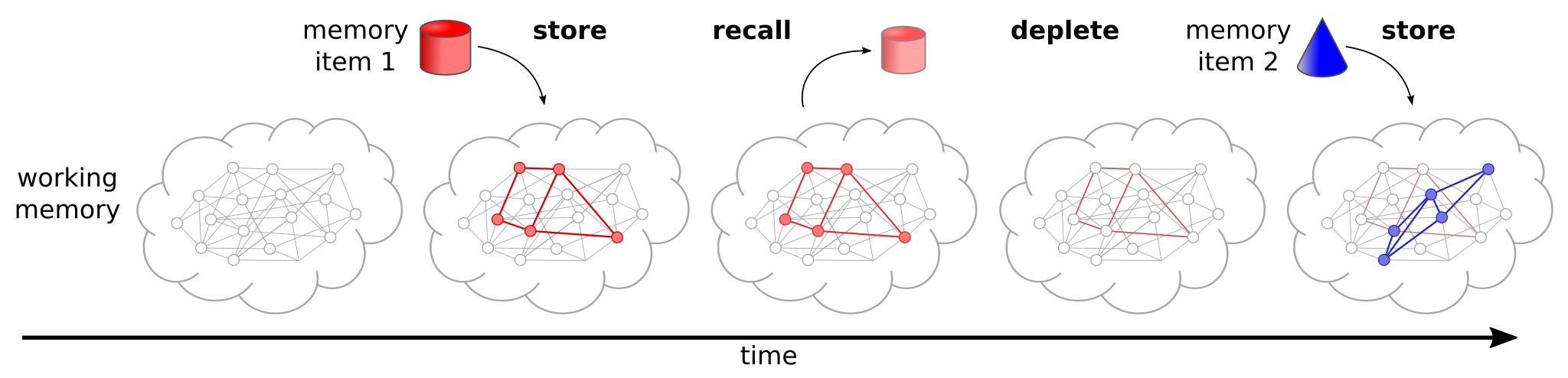}
    \caption{Conceptual illustration of information storage in working memory. The neural network can store and recall features of an item. The volatile nature of the synapses allows the memory of the stored features to fade in time. After depletion of the memory, the features of a different item can be stored without significant interferences.}
    \label{fig:fig1}
\end{figure}

The short-term storage of information, one of the main properties of \gls{wm}, is at the forefront of most cognitive abilities in humans and animals \cite{barak2014working, mi2017synaptic}. Tasks that involve \gls{wm} include visual processing, speech comprehension, and episodic planning; thus impairment of \gls{wm} results in a partial or complete loss of these abilities \cite{cowan2010magical,miller2018working}. Several models have been suggested to explain the physiological mechanisms that underlie the \gls{wm} in the brain. Today it is becoming increasingly evident that a mixture of specialized mechanisms enable an ensemble of versatile memory systems in the brain that cover memory duration from several seconds to minutes \cite{mongillo2008synaptic, miller2018working}.

One such mechanism for \gls{wm} is based on short-term dynamics of synapses, i.e., \gls{stp}. Figure~\ref{fig:fig1} illustrates the synaptic model of \gls{wm}. When a new memory item is stored in the \gls{wm} network, an ensemble of neurons that represents the item is activated and synaptic strengths between these neurons are transiently potentiated. This transient increase in synaptic strength is caused by short term mechanisms that put the synapse in a state of higher effectiveness \cite{mi2017synaptic, mongillo2008synaptic}. This state of high effectiveness supports the system to retrieve the stored memory item. 
The transient decay of the synaptic strengths and, thus, of the memory items makes the \gls{wm} susceptible for new content.

Various mechanisms for \gls{wm} were introduced also in modern \gls{ml} to improve the performance in complex sequence processing tasks including natural language \cite{weston2014memory, graves2014neural, vinyals2015pointer, santoro2016meta}. Indeed, similar to the \gls{wm}, \glspl{rnn} provide the possibility to perform associative memory and recall stored information thanks to the presence of internal feedback loops that ensures the persistence of data. In recent studies, recurrency demonstrated to be a crucial aspect in, e.g., grammar learning \cite{alamia2020SciRep}. In contrast to feedforward architectures that are often being used in \gls{ml}, \glspl{rnn} inhere the desirable computational properties to represent long-term dependencies in sequential data, due to their computational load \cite{dehghani2018universal, schafer2006recurrent}.

Network architectures such as \gls{lstm} and \glspl{gru} try to optimize the use of hardware resources while still exploiting recurrency \cite{yu2019NeurComp}. However, these solutions are still expensive in terms of hardware area and number of operations. Bespoke hardware optimizing resources in terms of power consumption, area, and computational workload could therefore facilitate the implementation of \glspl{rnn}. At present, \gls{cmos}-only solutions include digital standard hardware, e.g., \glspl{gpu} and \glspl{fpga}, as well as analog or mixed-signal \gls{asic}. Digital circuits usually imply a heavy computational workload and the need of waiting for several clock cycles \cite{amirsoleimani2020AdvIntellSys}, while \glspl{asic} have a higher energy efficient approach by adopting subthreshold circuits working asynchronously \cite{rahimi2020AdvIntellSys}. However, the temporal dynamics are usually implemented by charging or discharging a capacitor with a constant current \cite{chicca2014ProcIEEE}, thus consuming power. Furthermore, when dealing with biologically relevant time scales, the size of the capacitors becomes non-negligible.

Alternative approaches with better scaling and lower energy consumption are deeply sought to enable neuromorphic circuits with bio-realistic \glspl{wm} \cite{covi2021FrontNeurosci}. In this respect, a promising technology is represented by memristive devices \cite{christensen2022NCE}, namely two-terminal devices able to reversibly change their conductance upon the application of proper electrical stimuli. Memristive devices show a broad range of attractive properties, including high scalability, high read/program speed, high energy efficiency, and programming voltages comparable with the power supply of typical neuromorphic chips \cite{kuzum2013Nanotech,ielmini2019Nanotech}. While non-volatile memristive devices have already shown promising results when used as non-volatile synapses, volatile memristive devices that can keep track of recent neural activity and implement synapses with biologically compatible time constants, are still largely unexplored. Indeed, volatile properties have so far mainly been investigated in reservoir computing applications \cite{du2017NatComm,midya2019AdvIntellSys,zhong2021NatComm}.

In this work, we use C\,/\,HfO$\mathrm{_{2}}$\,/\,Ag memristive devices featuring a high ON/OFF ratio of 10$^8$ and tunable time constants in the range of milliseconds to seconds \cite{wang2021AdvIntellSys,covi2021TED-I}, which are two desirable features for our application. The switching mechanism relies on the formation and spontaneous dissolution of a silver \gls{cf} across the switching layer. The relaxation, namely retention time, i.e., the time it takes to the \gls{cf} to dissolve, has been shown to depend on the diameter of the \gls{cf} itself: the thicker the filament, the longer the retention time \cite{wang2019TED-I,wang2019TED-II,wang2021TED-II}. The advantages of this technology are manifold: the retention times are electrically tunable and the time information is physically located in the \gls{cf}, thus consuming a negligible power and area on the chip. Moreover, the probability to switch the device on is also electrically tunable, thus adding a further degree of freedom in controlling the dynamics of the \gls{wm}. These features are exploited to demonstrate store and recall of patterns in a small scale \gls{wm} hardware, then the system is scaled up in simulations to demonstrate the use of volatile memristive devices in a large-scale biologically inspired model of synaptic \gls{wm} and in an associative symbolic \gls{wm}.

\section{Results}
\subsection{Volatile memristive device}

The volatile synapse used in this work is a resistive switching C\,/\,HfO$\mathrm{_2}$\,/\,Ag memristive device (see fabrication details in \textit{Materials and Methods}). After the device fabrication, the device is in its high-resistive state and can be switched to the \gls{lrs} by applying a quasi-static voltage sweep between 0\,V and 1.5\,V and back. Contrary to many filamentary devices, the proposed device does not require an electroforming operation (Fig.~\ref{fig:Supfig_01}a), thus simplifying the electrical operation within the neuromorphic circuit. The maximum current flowing through the device (i.e., current compliance, I$\mathrm{_{CC}}$) is limited by applying a voltage to the gate of an NMOS transistor whose drain is connected to the bottom electrode of the memristive device (Fig.~\ref{fig:fig2}A, inset). The switching to the \gls{lrs} occurs because, as the voltage exceeds a given \textit{threshold voltage} V$\mathrm{_{T}}$, Ag ions migrate across the oxide layer and form the \gls{cf}, thus bringing the device in a \gls{lrs} (\textit{set operation}, Fig.~\ref{fig:fig2}A) \cite{Wang2016, Li2020, wang2019TED-II}. However, the filament self-sustains only in presence of a voltage higher than a critical voltage referred to as \textit{hold voltage} V$\mathrm{_{H}}$. Below this value, the rediffusion of the Ag atoms in the dielectric layer causes the self-disruption of the filament and, as a consequence, the device transition to a \gls{hrs} (\textit{spontaneous recovery}, Fig.~\ref{fig:fig2}A). These operations are fairly reproducible (Fig.~\ref{fig:Supfig_01}b). The two main features that enable the use of the proposed memristive device as a volatile synapse in working memory tasks are tunable volatility in biologically relevant time-scales and controllable switching probability.

The retention time is shown in Fig.~\ref{fig:fig2}B and the median retention time increases exponentially with the current compliance, as illustrated in Fig.~\ref{fig:fig2}C. The device can be tuned to be in \gls{lrs} for a time ranging from ms to seconds depending on I$\mathrm{_{CC}}$, which is the time-scale of interest for our application. The intrinsic stochasticity of the filament formation at a microscopic level originates high variability in the distributions of the retention times \cite{Chekol2021}.

The controllability of the switching probability (P$\mathrm{_{ON}}$) of the device can instead be achieved by either changing the width or the amplitude of the voltage pulse applied to the device, as shown in Fig.~\ref{fig:fig2}D.  
It is worth noticing that the device operates with voltages $<$3\,V, which are compatible with standard \gls{cmos} technology and therefore suitable for integration with neuromorphic circuits. Fig.~\ref{fig:fig2}E shows the stochastic behavior of the device. Due to its stochastic nature, the threshold voltage of the device is slightly variable, which results in an electrically tunable switching probability: The higher the pulse voltage, the higher the switching probability \cite{covi2021TED-I}. Therefore, the desired switching probability of the device can be tuned by setting a suitable voltage amplitude. Moreover, when stimulated by a burst of identical pulses (Fig.~\ref{fig:Supfig_02}), the switching probability of the device increases with the number of pulses, as demonstrated in Fig.~\ref{fig:fig2}F and \ref{fig:Supfig_03bis}. 
This effect can therefore be exploited to accelerate the storing or training phase in a network.

\begin{figure}[ht]
    \centering
    \includegraphics[width=0.95\textwidth]{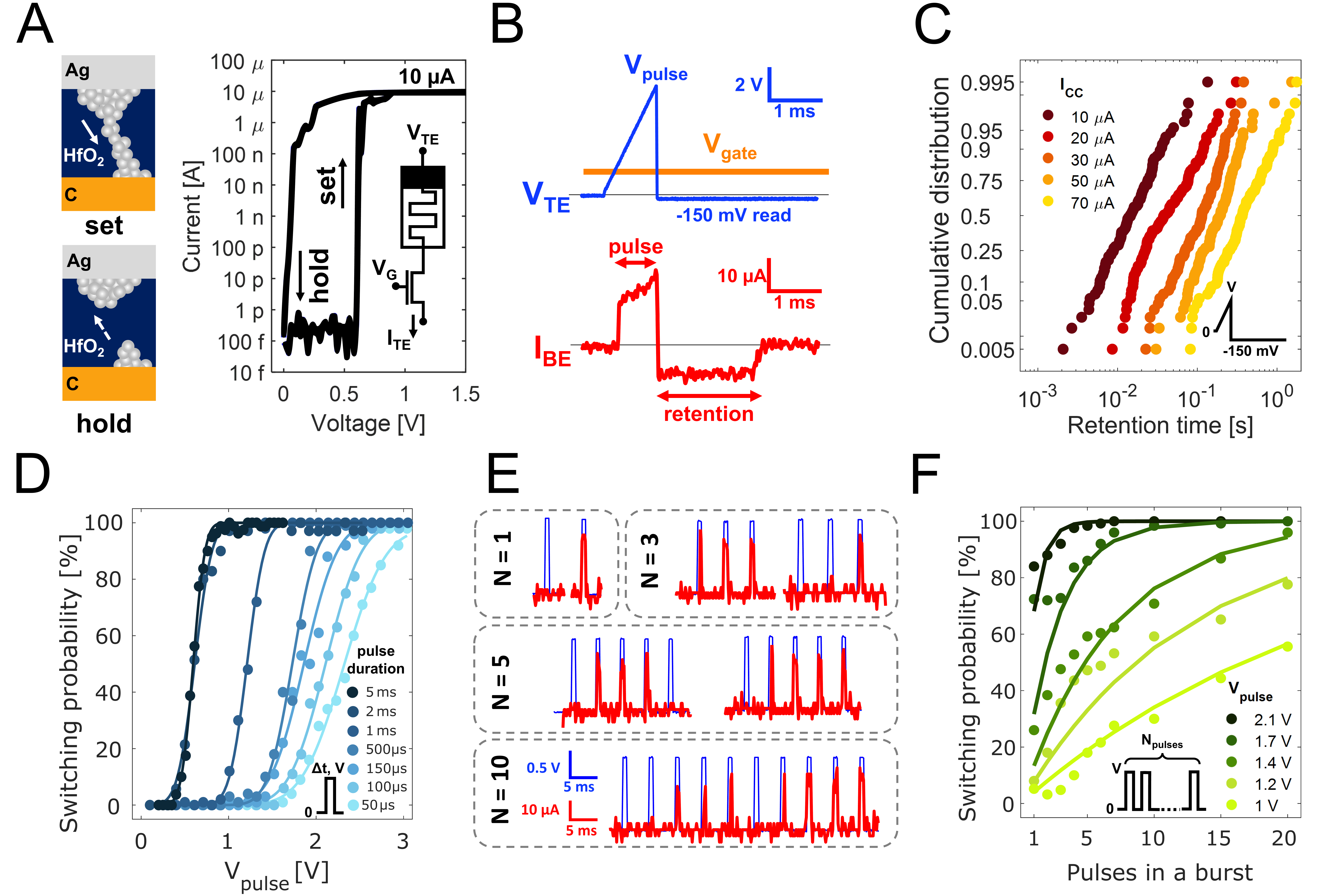}
    \caption{Ag-based volatile memristive device characterization. (a) Sketch of the one-transistor\,/\,one-resistor (1T1R) RRAM device together with its working principle. The memristive device (1R) is based on a W\,/\,C\,/\,10\,nm HfO$\mathrm{_2}$\,/\,Ag stack. The RRAM shows a volatile behavior, i.e., a set operation together with a spontaneous switch off. (b) Time characterization of the retention of the filament. After a 5\,V amplitude triangular pulse to switch  the cell on with an I$\mathrm{_{CC}}$\,=\,20\,\textmu A, a constant reading voltage of -150\,mV is applied to monitor the retention. (c) Retention time distributions at different compliance currents. The median value of the retention time increases with the compliance current. Inset: applied programming pulse. (d) Switching probability of the device for a single pulse as a function of the amplitude and the pulse duration of the programming pulse. Shorter pulses required higher voltage amplitudes to switch ON. (e) Probability of switching the RRAM to the \gls{lrs} depending on the number of pulses and their amplitude (1\,ms pulses). The circles are the experimental data while the solid lines is the fitting. (f) Effect of the number of pulses and the voltage amplitude (1.6\,V) on the switching of the device. The switching of the device is stochastic. Considering a group (burst) of pulses, the probability that the device is in the \gls{lrs} inside the group increases.}
    \label{fig:fig2}
\end{figure}

\subsection{Working memory}
\subsubsection{Store and recall of features}

The volatile behavior of the memristive device is used in a small-scale experiment of \gls{wm} to store and recall features, as depicted in Fig.~\ref{fig:fig3}A. In our example, our network consists of 5 volatile synapses all connected to the same neuron (schematic in Fig.~\ref{fig:fig3}B). The current flowing to the neuron is the sum of the currents flowing through the stimulated devices and it depends on the resistive state of each device. The neuron is trained to recognize a color within a color stream. Each color is encoded by stimulating a different combination of three devices as in Fig.~\ref{fig:fig3}C. As a result, for each color only three devices out of five are switched on. At first, the target color is stored in the network by repeatedly stimulating the network with the combination of pulses that encodes the desired color. As a consequence, the corresponding stimulated devices will be switched on and set to their \gls{lrs}. Afterwards the network is fed with a stream of random stimuli, as shown in Fig.~\ref{fig:fig3}D (see also Fig.~\ref{fig:Supfig_03}). When the stored item is presented to the network, all the stimulated devices are in \gls{lrs} and therefore the current received by the neuron overcomes a threshold, thus triggering the firing of the neuron. The current threshold is set according to the current distributions shown in Fig.~\ref{fig:Supfig_04}. After about 1\,s without stimulation, all the devices switch off and the pattern is forgotten, as shown in Fig.~\ref{fig:Supfig_03}. The correlation plot in Fig.~\ref{fig:fig3}E shows the difference between the expected and the measured current when a pattern is presented. The main contribution to the current is given by the synapses that are stimulated and in \gls{lrs}, therefore three different current levels can be expected. It is shown that we can reach more than 90\% accuracy, defined as the percentage of correct classifications (stored / non-stored pattern) during a test sequence. 

\begin{figure}[ht!]
    \centering
    \includegraphics[width=0.95\textwidth]{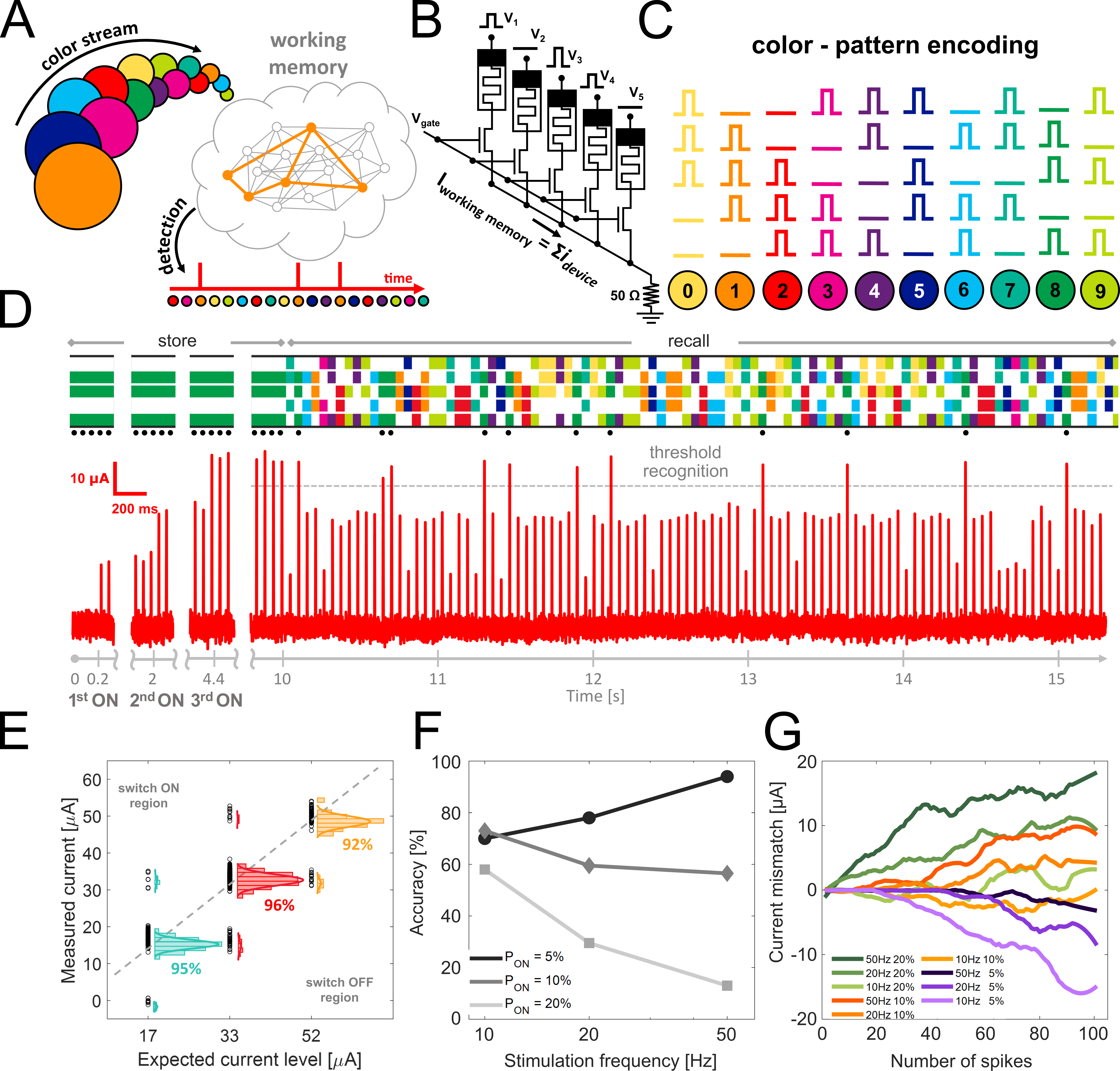}
    \caption{\Gls{wm} store and recall experiment. (a) High-level sketch of the working memory. (b) Schematic of the \gls{wm} implementation: 5 volatile 1T1R memristive devices are arranged in parallel configuration. The gate is chosen to set I$\mathrm{_{CC}}$\,=\,17\,\textmu A, that corresponds to a retention time of 28\,ms. (c) Color - pattern encoding. (d) Store and recall experiment. During the store phase, a single pattern is fed to the network. Top colored plot: input stimuli. For ease of visualization, each pattern is colored as the color it encodes. Black dots in the bottom part of the upper plot indicate the stored pattern. Bottom plot: measure current fed to the post-neuron. The current threshold for recognition is indicated as a dashed black horizontal line. The traces are cropped on the x-axes to better highlight the salient events. (e) Correlation plot between the expected and measured currents based on the difference between the presented and the stored pattern. Results obtained from 10 different store and recall experiments with P$\mathrm{_{ON}}$\,=\,5\% and stimulation frequency f$\mathrm{_{stim}}$\,=\,50\,Hz. (f) Accuracy of the system in distinguishing the stored pattern under different stimulation and switching conditions. (g) Average current error, defined as the difference between the measured current and the expected current, during 100 patterns applied for the different conditions. 
    }
    \label{fig:fig3}
\end{figure}

The retention time and the switching probability of each synaptic device are electrically tunable through the gate voltage of the transistor (which sets the current compliance) and the voltage amplitude of the pulse applied to the \gls{te} of the device, respectively. These features, together with the stimulation frequency of the network, determine the accuracy of the network in recognizing the color. As shown in Fig.~\ref{fig:fig3}F, for low P$\mathrm{_{ON}}$ (P$\mathrm{_{ON}}$\,=\,5\%), an increase of the spike rate leads to an improvement of the accuracy because the time between two pulses is shorter than the retention time of the device. Instead, higher values of P$\mathrm{_{ON}}$ lead the network to experience a decrease in its accuracy because the probability that a non-active device is switched on is higher. Finally, the intrinsic volatility of the network results in a progressive forgetting of the stored element. Fig.~\ref{fig:fig3}G gives an estimation on how long the network remembers by showing the average mismatch current, defined as the difference between the expected current and the measured current. The two main errors that can occur over time, i.e., the network either fails to recognize the correct color or recognizes the wrong one, depend on the combination of stimulation frequency and P$\mathrm{_{ON}}$. Indeed, for high P$\mathrm{_{ON}}$ and stimulation frequency, devices that were supposed to be in their \gls{hrs} turn to their \gls{lrs}, thus increasing the measured current. The opposite combination, i.e., low P$\mathrm{_{ON}}$ and stimulation frequency, results in the switching off of devices previously in \gls{lrs}, thus decreasing the measured current. The results suggest that indeed a careful selection of the P$\mathrm{_{ON}}$ based on the planned stimulation frequency is advisable to fine tune the \gls{wm} accuracy for a given task, as it will be discussed further in Section~\ref{sec:disc}.

\subsubsection{Biologically inspired model of synaptic \gls{wm}}

\begin{figure}[ht]
    \centering
    \includegraphics[width=\textwidth]{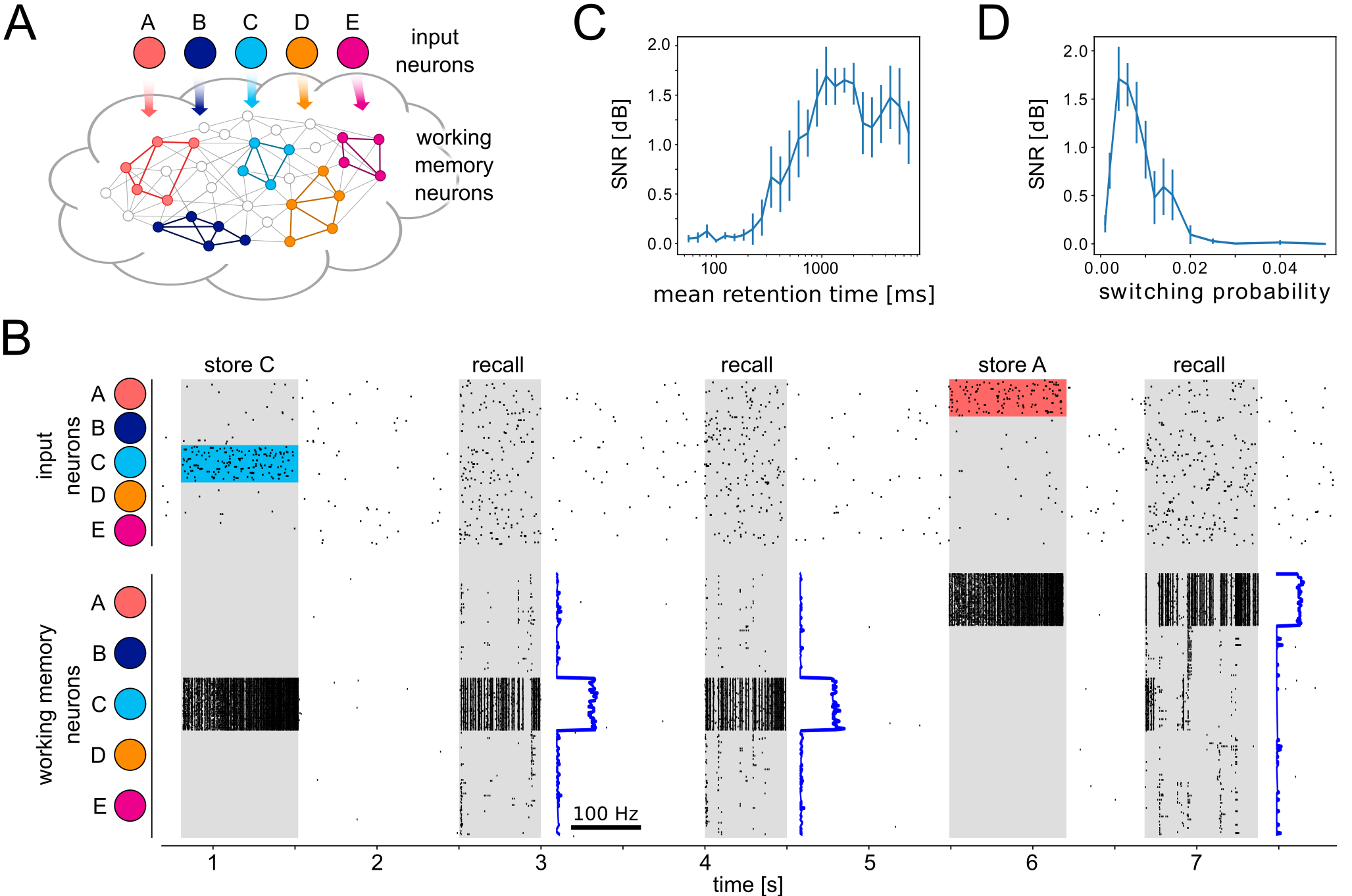}
    \caption{Large-scale simulation of \gls{wm}. (a) Illustration of the network model. 5 different memory items (\textit{A,B,C,D} and \textit{E}) can be stored in a recurrent network of spiking neurons. Corresponding strongly connected populations within the network transiently store memory items after activation.
    (b) Network activity of the \gls{wm} model. Black dots show individual spikes of input (top) and network (bottom) neurons. Multiple phases of store and recall are shown. Insets show average firing rates (spikes per second in Hz) over recall phases. Data obtained using a current compliance of 330\,\textmu A, corresponding to average retention times of 1.5\,s, and a voltage amplitude of 0.5\,V, corresponding to a switching probability of 5\%, were used in this simulation.
    Network behavior using (c) different retention time distribution (change of compliance current) and (d) different switching probability (change of applied voltage amplitude).}
    \label{fig:tsodyks-wm}
\end{figure}

The experimental results on \gls{wm} based on volatile memristive devices can be used to develop a \gls{wm} model for simulation of a larger scale, recurrent neural network. In \cite{mongillo2008synaptic} a model of \gls{wm} in the brain is introduced that is based on \gls{stp} combined with multi-stable network dynamics. We adapted this network model to implement the core dynamics of our volatile memristive synapses (Fig.~\ref{fig:fig2} and \ref{fig:fig3}) as a model of \gls{stp}. Fig.~\ref{fig:tsodyks-wm}A illustrates our model that serves as \gls{wm} model that is able to store a set of discrete patterns. For this, the network receives input from five input populations that represent the input patterns to be stored. 
A set of input neurons (illustrated by colored circles) are connected to the recurrent \gls{wm} network (cloud) so that the memory items could be stored and reactivated. All neurons are spiking neurons, such that outputs are given by unitary events that are emitted when an internal membrane potential variable crossed the firing threshold (see Section~\ref{sec:network}).
The network has a multi-stable dynamics in that neurons of a specific population excite each other and transiently strengthen synapses within the population through \gls{stp} when activated. Inhibitory neurons are added to facilitate the network multi-stability (see Section \textit{Materials and Methods} for details). The experimental paradigm, the network architecture, and the synapse parameters are adjusted to the characteristics of our memristive devices. Retention time parameters are fit to the measured device characteristics.

Fig.~\ref{fig:tsodyks-wm}B shows typical behavior of the model over a simulation of several seconds. Spiking activity of the network and input neurons are shown. The pattern \textit{C} is stored by strongly activating the corresponding input neurons. This triggers co-activation of the pool of \gls{wm} neurons that encode patterns \textit{C}. Through this activation, volatile memristive synapses get strengthened and cause a prominent response in \gls{wm} neurons when a recall stimulus that activates all input neurons at intermediate rates is given after a timeout of 1\,s. During this recall phase neurons that encode pattern \textit{C} show significantly increased activity. Two repeated recalls are shown and lead to reliable memory performance. In the timeout phase \gls{wm} neurons are almost perfectly silent, which is perfectly in line with the behavior of biological neurons. After the second recall phase the memory item \textit{C} is being forgotten and \textit{A} can be stored without interference by strongly activating pattern \textit{A} input neurons. In a third recall period, \textit{A} but not \textit{C} neurons get strongly activated.

The \gls{wm} model requires time constants in the order of behaviorally relevant time scales ($>$500\,ms to few seconds) to function well. Fig.~\ref{fig:tsodyks-wm}C shows the  memory recall performance of the network when different retention times are used. Recall performance is measured here as the \gls{snr} between the \gls{wm} neurons. The \gls{snr} is computed here as the mean firing rate of the population that is specific to the memory item over the mean firing rate of non-specific populations. Retention times below 1\,s lead to degraded performance because the firing activity of the neurons is significantly lower than the retention time. This result is in accordance with the experiments shown in Fig.~\ref{fig:fig3}. Also switching probabilities had to be finely tuned. Fig.~\ref{fig:tsodyks-wm}D shows the impact of switching probabilities on the recall performance. Switching probabilities of about 0.05 are found to work best for this memory store/recall protocol because it prevents an excessive activation of the volatile devices, which would prevent the correct storage of information, as also confirmed by the experimental results of Fig.~\ref{fig:fig3}. Thanks to the flexibility of our volatile memristive devices, these probabilities can be obtained with a suitable selection of the voltage amplitude and pulse time width as shown in Fig.~\ref{fig:fig2}D.

\subsubsection{Associative symbolic \gls{wm}}

\begin{figure}[ht]
    \centering
    \includegraphics[width=\textwidth]{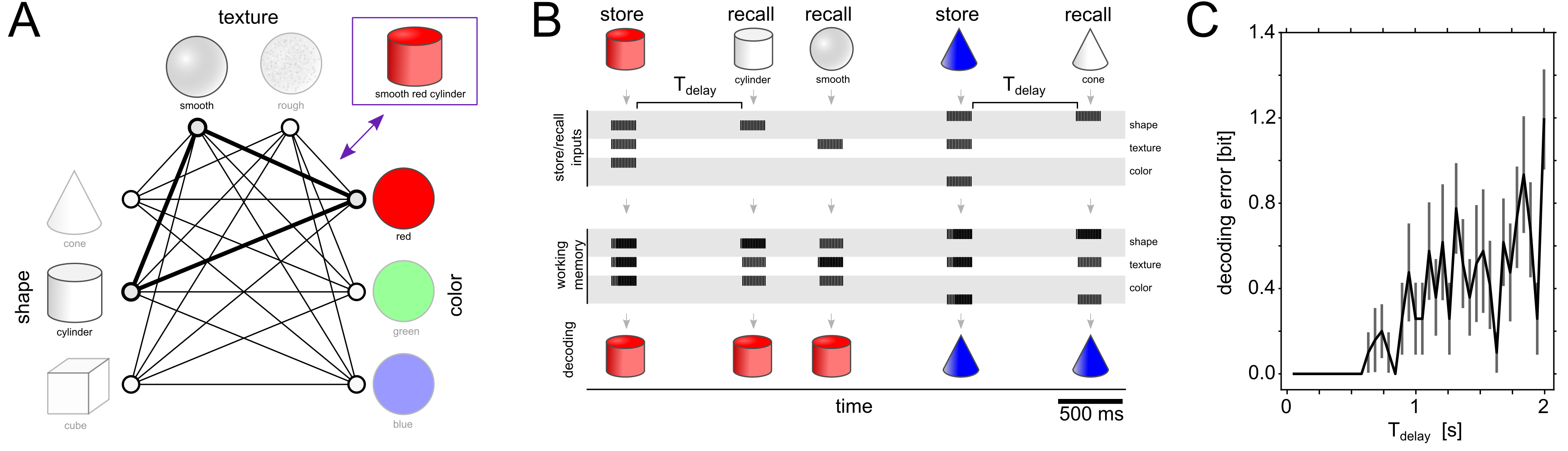}
    \caption{Associative symbolic \gls{wm} with volatile memristive devices. (a) Illustration of the network for associative symbolic \gls{wm}. (b) Sequence of store and recall. Store/recall input (top row) and \gls{wm} neuron output spikes (bottom row) are shown. Inserted pictograms represent the decoded objects and recall queries. Store/recall inputs had a one-to-one fixed connectivity to \gls{wm} neurons. After storing an association it can be recalled by cuing the network with an arbitrary memory element. (c) Decoding error plotted as a function of time delay between store and first recall.}
    \label{fig:symbolic-working-memory}
\end{figure}

Another important feature of \gls{wm} is the ability to transiently form associations between properties, such as color or shape, to remember representations of real-world object. Fig.~\ref{fig:symbolic-working-memory}A illustrates this associative symbolic \gls{wm} model. 
We investigated whether the model memristive devices can be used to form transient associations between symbolic items of different categories. We used a similar store-recall paradigm as in Fig.~\ref{fig:tsodyks-wm} but here the memory item to store is given by association between features that could be dynamically bound together. These features are thought to represent the state of physical objects an associative memory networks is able to perceive through a set of sensors. Concretely we used features from 3 different stimulus categories, shape, texture and color (see Fig.~\ref{fig:symbolic-working-memory}). Objects are represented by jointly activating feature neurons in the memory network, e.g. a \textit{`smooth, red cylinder'} is represented by jointly activating the corresponding feature neurons in the network. STP model synapses as in Fig.~\ref{fig:tsodyks-wm} are used as short-term memory inside the bidirectional connections between the neurons.

The network is able to form an autoassociative memory. After an association has been formed by presenting an object to the network, the features of that object can be recalled, after a brief time delay of length $T_\text{delay}$, by querying the network with only one part of the features. Fig.~\ref{fig:tsodyks-wm}B shows one example store and recall sequence. After the object \textit{`smooth, red cylinder'} is stored, neurons representing its features can be re-activated by triggering only the \textit{`cylinder'} or \textit{`smooth'} neurons. After a new object (\textit{`smooth, blue cone'}) is stored, its representation can be retrieved by only activating the \textit{`cone'} neuron. The object representations are transiently stored for and then slowly fade away. This is analyzed in Fig.~\ref{fig:tsodyks-wm}C where we plot the decoding error as a function $T_\text{delay}$. The memory can be reliably retrieved during a time window of 600\,ms, corresponding in our devices to I$\mathrm{_{CC}}$\,=\,70\,\textmu A.

\section{Discussion}
\label{sec:disc}

The emergence of a new class of volatile memristive devices featuring volatility in biologically-relevant time scales opens the possibility to implement in hardware neuromorphic systems that can inherently solve complex sequence processing tasks. The main advantage of the proposed technology lies in the storage of the information in the physical configuration of the nanoscale device, i.e., the \gls{cf} inside the oxide layer of the memristive device. This way, a direct correlation of the retention time with the electrically tunable properties of the device is established.

The volatile properties of memristive devices have been so far exploited mainly in reservoir computing \cite{du2017NatComm,midya2019AdvIntellSys,zhong2021NatComm}. Other applications include selector devices and hardware security \cite{wang2020AdvIntellSys}, while the exploration of the potential of volatile devices in systems requiring short-term memory is still at its infancy \cite{ji2022AdvIntellSys,giotis2022palimpsest}. Yet, the use of volatile devices in tasks as \gls{wm} is extremely advantageous from a hardware perspective. Indeed, the networks devoted to carry out such tasks need the ability to forget the stored information in time, otherwise the network would quickly reach its maximum memory capacity and become unable to store new experiences unless old ones are forgotten. An implementation using non-volatile devices is thus feasible only together with the design of extra circuits that could reset the devices, thus consuming extra area and power. Another solution to implement volatility is with capacitors. However, in addition to the much larger area that capacitors would require to implement the same time constants as the proposed volatile devices, the stochastic properties that contribute to the correct functioning of the network (see Fig.~\ref{fig:tsodyks-wm}D) would have to be implemented by dedicated circuits.

In our work, we exploit the properties of Ag-based resistive switching devices to carry out \gls{wm} tasks. We first characterize the device and assess its electrical properties, then we select the parameters to conduct a store and recall hardware experiment, where we demonstrate the ability of short-term storage of features and explore the performance of the system under a variety of stimulation conditions.
We then use a biologically inspired synapse model that qualitatively reproduces short-term dynamics of \gls{stp} with the properties of our memristive device \cite{zucker1989short, tsodyks1998neural}. We demonstrate the functionality of this \gls{stp} model by qualitatively reproducing the \gls{wm} model of \cite{mongillo2008synaptic} in Fig.~\ref{fig:tsodyks-wm}.
This experiment demonstrates that \gls{stp} based on volatile memristive device dynamics is able to install \gls{wm} capability in a multi-stable network of spiking neurons. Memory items can be reliably retrieved seconds after storage and overwritten with new items on the same time scale. Furthermore we demonstrate a symbolic associative working memory model in Fig.~\ref{fig:symbolic-working-memory} where we use the data of the device characterization to preserve the feature of the device in terms of switching probability, retention time, and variability.

As shown in Fig.~\ref{fig:fig3}, the device parameters and the stimulation conditions are linked and therefore they need to be matched in order to achieve successful network operation. During the experiments, the store phase exploits the properties of burst stimulation shown in Fig.~\ref{fig:fig2}E, which in our case is beneficial because it shortens the store phase. Due to their stochasticity, the devices do not switch on together, as visible in Fig.~\ref{fig:fig3}D, hence the duration of the store phase should take this aspect into consideration. In case of two consecutive store phases, only the second element is actually safely stored in the network. Indeed, during the second store phase, the synapses common to both elements remain active, whereas the ones that are no longer stimulated switch off, while the others specific to the second element only switch on.
Changing P$\mathrm{_{ON}}$ has repercussions in the speed of the store process (Fig.~\ref{fig:Supfig_05}), but also in the recall phase (Fig.~\ref{fig:fig3}F and Fig.~\ref{fig:fig3}G). Indeed, while an increase of P$\mathrm{_{ON}}$ is beneficial during store, it becomes detrimental during recall, since it results in the erroneous activation of one or more devices. A trade-off therefore exists between store speed and recall accuracy when setting the device's P$\mathrm{_{ON}}$.

Another important factor that has to be considered when setting the device parameters is the expected stimulation rate. Each incoming pulse has a different effect on the memristive device, depending on whether the device is ON (in \gls{lrs}) or OFF (in \gls{hrs}). In the former case, the \gls{cf}, i.e., the information stored in the memristive device, is refreshed. In the latter, the stimulation might activate OFF devices. As a consequence, there are two mechanisms that degrade the accuracy of the system: Low stimulation rates may lead to the switching off of previously ON devices, while high rates may lead to erroneous classification due to the switching on of previously OFF devices (Fig.~\ref{fig:Supfig_06}, Fig.~\ref{fig:Supfig_07}). A possible mitigation measure is tuning the mean retention time of the devices: low spike rates need longer retention times, whereas high spike rates might benefit from shorter retention times.

The proposed \gls{stp} synapse model presents an important difference compared to the one in \cite{zucker1989short, tsodyks1998neural}, that is the high level of noise, which imitates a feature of biological synapses. 
Chemical synaptic transmission in biology is inherently unreliable. About half of synaptic transmissions are not detectable at the post-synaptic side at all, which makes synapses an abundant source of noise in the brain \cite{rusakov2020noisy, borst2010low, jensen2019multiplex}. Given how costly synapses are in terms of energy consumption \cite{pulido2021synaptic}, this finding is surprising. Several authors have therefore suggested that the noise in synapses serves as a computational resource that allows the brain to solve complex tasks more efficiently \cite{rusakov2020noisy, maass2014noise}. We show here that the noise in synaptic short-term dynamics, that mimics the behavior of synaptic facilitation in biological synapses, can be exploited to realize short-term memory on behaviorally relevant time scales of several seconds.

\section{Conclusion}
\label{sec:concl}
In summary, real-world applications require very different time constants. Technologies that enable the design of systems whose internal temporal dynamics can be tuned to match the real world ones present appealing opportunities, especially in the context of power and memory limited edge computing. Our results show that the proposed Ag-based volatile memristive device features electrical tunability of its key parameters, i.e., retention time and switching probability, that allows to adapt the lifetime of \gls{wm} to the task-specific timescale needed - from 1\,ms to 10\,s.

\section{Materials and Methods}

\subsection{Device fabrication}
\label{ssec:dev_fab}
The memristive devices used in this study are fabricated on top of foundry-based MOS transistors, namely one-transistor~/~one-resistor (1T-1R), allowing the control of the compliance current $I_\mathrm{C}$~\cite{bricalli2018TED-I}. The \gls{be} consists of a 70\,nm\,$\times$\,70\,nm graphitic carbon pillar, that was already demonstrated to be a good electrode material for both volatile and non-volatile devices thanks to its stability and inert behavior \cite{covi2021TED-I}. The oxide layer and the \gls{te} are fabricated by e-beam evaporation. The oxide is a 10\,nm HfO$\mathrm{_2}$ active layer and the \gls{te} is a 100\,nm thick Ag layer. Both HfO$\mathrm{_2}$ and Ag are deposited at room temperature and without breaking the vacuum (pressure 3$\times$10$\mathrm{^{-6}}$\,mbar).

\subsection{Electrical setup for device characterization}

The electrical setup is the same as described in \cite{covi2021TED-I}, where the device is connected either to a semiconductor device parameter analyzer (DC characterization) or to a waveform generator and an oscilloscope (pulsed characterization). DC characterization is carried out using a HP~4156C semiconductor device parameter analyzer. A sweep from 0\,V to 1.5\,V is applied to the \gls{te} of the device while the \gls{be} is grounded. The characterization includes DC sweeps with different compliance current (I$\mathrm{_C}$), as in Fig.~\ref{fig:Supfig_01}a. The pulsed characterization as well as the working memory experiment are carried out using a TTI~TGA12104 arbitrary waveform generator. The voltage waveforms are applied to the \gls{te}. To measure the current, a LeCroy Waverunner 640Zi oscilloscope is connected at the \gls{be} side, and the voltage drop across a 50\,$\Omega$ series resistance is probed. MATLAB\textsuperscript{\textregistered} software is used for the data analysis and the control of the instruments.

The temporal dynamics are studied by first applying a semi-triangular pulse with 10\,ms pulse duration and 5\,V amplitude to induce the filament formation and then monitoring the state of the filament with constant -150\,mV bias. To tune the temporal behavior, the I$\mathrm{_C}$ is changed from 10\,\textmu A to 70\,\textmu A. For each I$\mathrm{_C}$, 100 experiments are carried out. To avoid possible interference due to the previous cycles, the value of I$\mathrm{_C}$ is changed in random order.

The switching probability (Fig.\ref{fig:fig2}D) is studied by applying 100 pulses for each combination of voltage amplitude - pulse duration. The conditions are applied in random order. The impact of the number of pulses (Fig.\ref{fig:fig2}D and E) is analyzed with the same methodology, selecting the order of the combination of voltage amplitude - number of pulses randomly. Each combination is applied 100 times.

For the \gls{wm} store and recall experiment, each probability-frequency-pattern condition is repeated 10 times. The voltage bias is applied only at the end of the experiments to check the retention capabilities.

\subsection{Fitting of device features}

Thanks to the stochasticity of the switching mechanism, the probability P$_{ON}$ for a single pulse of given amplitude (Fig.~\ref{fig:fig2}D) follows a normal distribution and thus is fit using a cumulative distribution function:

\begin{equation} 
\label{eqPON}
{\Large\displaystyle{P_{ON}(V) = \frac{1}{2}\left(1+erf\left(\frac{V-\mu}{\sigma\sqrt{2}}\right)\right)}}
\end{equation}

where the average value $\mu$ and variance $\sigma$ are fitting parameters of the device and the pulse duration, respectively. Table~\ref{tab:tab1} collects the data of the device presented in Fig.~\ref{fig:fig2}D. Each point in the figure corresponds to 100 measurements. 

\begin{table}[ht]
    \centering
    \caption{Values of $\mu$ and $\sigma$ calculated for different pulse time widths.}
    \begin{tabular}{|c|c|c|}
        \hline
        \textbf{Time width [ms]} & \textbf{Average $\mu$ [V]} & \textbf{Variance $\sigma$ [V]} \\
        \hline
        0.05 & 2.31 & 0.38 \\
        \hline
        0.10 & 2.11 & 0.33 \\
        \hline
        0.15 & 1.86 & 0.30 \\
        \hline
        0.50 & 1.73 & 0.22 \\
        \hline
        1.00 & 1.21 & 0.16 \\
        \hline
        2.00 & 0.61 & 0.15 \\
        \hline
        5.00 & 0.59 & 0.11 \\
        \hline
    \end{tabular}
    \label{tab:tab1}
\end{table}

The switching probability, for a given amplitude, as a function of the number of pulses is fit with the mathematical model: 

\begin{equation} 
\label{eqPON_N}
P_{ON}(N) = 1 - (1 - P_{ON}(1))^N\quad \mathrm{where} \quad P_{ON}(1) = P_{ON}(V)
\end{equation}

In the \gls{wm} store and recall experiments, unless otherwise noted, the following conditions apply: no read voltage between spikes during the store and the recall phases. A read voltage of -150\,mV is applied at the end of the experiment. The current compliance is I$\mathrm{_C}$ = 17\,\textmu A, which gives an average retention time of 28\,ms.

\subsection{STP synapse model}
\label{sec:stpwmm}

To perform the computer simulations we developed a simplified phenomenological model that qualitatively reproduces the the behavior of the memristive devices. The model captures the switching probabilities and variable retention time of the devices. Each synapse $i$ was modeled with a binary internal state variable $x_i$ that denotes either the low ($x_i=1$) or the high resistance state ($x_i=0)$. The device resistance was $r_1$ and $r_0$ in the low and high resistance state, respectively. To model switching probabilities we assigned a parameter $\rho_i \in [0,1]$ to every synapse. Upon arrival of a pre-synaptic input spike, $x_i$ was set to 1 with probability $\rho_i$.

To model the trial-by-trail variability of the retention times we adopted a Lognormal distribution for the retention times $t_{ret}$, namely:
\begin{equation}
  \begin{split}
    t_{ret} \; &\sim\; \mathrm{Lognormal}( t_{ret} \;|\; \mu_i, \sigma_i )
    =\; \frac{1}{\sqrt{2}\pi\sigma_i \, t_{ret}} \, \exp \left( -\frac{(\log(t_{ret}) - \mu_i)^2}{2\sigma_i^2} \right) \;.
  \end{split}
  \label{eqn:lognormal}
\end{equation}
The parameters $\mu_i$ and $\sigma_i$ in Eq.~\ref{eqn:lognormal} were adjusted to fit the device properties. Supplementary figure~\ref{fig:Supfig_lognormal_fit} shows example model fits to experimental data for different compliance currents. In the simulations we used $\mu =  7.24$ and $\sigma_i = 0.82$, if not stated otherwise, which corresponds to a mean retention time of $\sim1.5$\,s. Whenever a synapse was set to the low resistance state ($x_i=1$) the retention time $t_{ret}$ was  drawn from Eq.~\ref{eqn:lognormal}. After the simulation time $t$ exceeded $t+t_{ret}$ the synapse spontaneously returned to the high-resistance state.

\subsection{Working memory network model}
\label{sec:network}

To reproduce the working memory model introduced in \cite{mongillo2008synaptic}, we used a recurrent network of 8000 excitatory and 2000 inhibitory neurons. Connection probabilities between these neurons were as in \cite{mongillo2008synaptic}. Supplementary Figure~\ref{fig:Supfig_08} illustrates the detailed network structure, connection probabilities and baseline synaptic conductances. Each memory item was represented by a population of 800 excitatory neurons, which were randomly chosen from the recurrent network for each of the 5 memory items. Synaptic weights between these neurons were 5 times stronger (0.5\,mV) than between other neurons (0.1\,mV). All inhibitory synapses were static (no STP) and had a strength of -0.2\,mV. To store and recall the working memory, 1000 input neurons were set up for each memory item. Input neurons were only connected to one of the memory item populations in the network. Input neurons fired at low baseline firing rates of 0.1\,Hz. To store a memory item, firing rates of input neurons corresponding to one memory item were elevated 10-fold. During recall all input neurons were set to unspecific elevated activation.

All neurons used the leaky integrate and fire (LIF) model with biologically plausible parameters \cite{gerstner2014neuronal}. The LIF neuron is a spiking point neuron model that transiently integrates synaptic inputs using a leaky membrane potential. The membrane potential $u(t)$ follows the dynamics
\begin{equation}
\frac{d\,u}{d\,t}  \;=\; - \frac{1}{\tau_m} \left(u(t) - u_0 \right) \;+\; i(t)\;,
\label{eq:membrane-potential}
\end{equation}
where $u_0$ is the membrane resting potential $\tau_m$ is a time constant. $i(t)$ is the input into the neuron denoting the summed effect of afferent synapses. If the membrane potential crosses a firing threshold $\vartheta$ at time $t^f$ a spike is emitted and the membrane potential is reset immediately after
\begin{equation}
 u(t^f+\Delta t) = u_r\;.
 \label{eq:reset_behavior}
\end{equation}
After a spike the neuron is inactive for a brief refractory time. The firing threshold was set to 20\,mV and refractory time to 2\,ms. Membrane time constants $\tau_m$ were 15\,ms and 10\,ms, resting potentials $u_0$ 16~mV and 13\,mV, for excitatory and inhibitory neurons, respectively. Independent unit variance Gaussian noise with mean 0.5775\,mV were injected into each excitatory, and with mean 0.5275\,mV into inhibitory neurons. As in \cite{mongillo2008synaptic} the mean of the Gaussian noise was precisely tuned to enable multistability in the network. Example network dynamics are shown in Fig.~\ref{fig:tsodyks-wm}D, for a store-recall experiment over 8 seconds.

\subsection{Associative symbolic working memory model}
\label{sec:associative-memory}

For the associative working memory model in Fig.~\ref{fig:symbolic-working-memory} we developed a network architecture that was capable to store bidirectional short-term associations between items of different categories. The associative memory model consisted of 8 memory neurons that were connected through bidirectional associative connections as illustrated in Fig.~\ref{fig:symbolic-working-memory}A. The memory neurons encoded for the three memory categories, shape, texture and color. Each memory neuron exclusively received input from one input neuron to trigger store and recall events. Memory neurons were implemented using the LIF neuron model, with firing threshold of 1.4\,V and membrane time constant of 15\,ms. Each group of neurons corresponding to one memory category was augmented with a single inhibitory neuron that provided lateral feedback to enforce mutually exclusive activation of one memory item at a time. Inhibitory neurons were LIF neurons with threshold of 1.4\,V and membrane time constant of 10\,ms. All excitatory connections had a strength of 1.5\,V and inhibitory feedback was -1.5\,V.

Model memristive devices were used inside the bidirectional connections to store the association between a specific pair of memory neurons (which we call here the specific neurons). Exactly one model device was used per association. Memristive devices were augmented with two auxiliary threshold gates to route the input and outputs during store and recall cycles. Gates were put in series with the devices, one before (input gate) and one after (output gate). Input gates received excitatory input from the two specific memory neurons the association corresponded to. In addition input gates received inhibitory input from all other memory neurons. Output gates recursively connected back to specific memory neurons through excitatory connections. Storage of memory items was not dependent on the activity of the neurons and threshold gates (see Fig.~\ref{fig:symbolic-working-memory}), but retained solely in the hidden state of the memristive devices. Model parameters of memristive devices were fit to experimental data ($\mu$=7.24, $\sigma$=0.82, corresponding to ca. 1.4\,s mean retention time). Switching probabilities $\rho$ were 0.2. The synaptic conductance for the low resistance state was chosen to elicit a voltage pulse of $1.0\,V$ in the post-synaptic neuron. The high resistance $r_0$ was set to be 20 times higher than the low resistance $r_1$.

\subsection{Details to software simulations}

Simulations of the working memory model with volatile memristive devices were done in python (3.8) using a custom implementation of the memristive synapse model that was developed for the NEST~2.14 simulation environment \cite{Peyser:838729}. The simulation time step was 1\,ms. Neuron dynamics were simulated using the current-based leaky integrate-and-fire model is available in NEST. A custom model of the STP model outlined in Section~\ref{sec:stpwmm} was implemented based on existing synapse models. Data analysis used the numpy and matplotlib python packages in version 1.23.2 and 3.5.3, respectively. The code will be made available upon publication.

\endgroup

\bibliographystyle{naturemag}
\bibliography{refs}

\section*{Acknowledgments}
The Authors thank Matteo Farronato and Alessandro Milozzi for fruitful discussions on the volatile devices. The Authors would like to thank Polifab's staff Marco Asa, Andrea Scaccabarozzi, Claudio Somaschini, Chiara Nava, Stefano Fasoli, Stefano Bigoni, and Elisa Sogne for help in the fabrication process. Thanks to Luciano Feltri for the help in the setup optimization.

\section*{Funding}
This work was partially supported by the European Research Council (ERC) through the European's Union Horizon Europe Research and Innovation Program under Grant Agreement No 101042585 and from the European Union’s Horizon 2020 Future and Emerging Technologies Program, Grant Agreement No 824164. Views and opinions expressed are however those of the Authors only and do not necessarily reflect those of the European Union or the European Research Council. Neither the European Union nor the granting authority can be held responsible for them.  C.T. acknowledges funding by the German Research Foundation (CRC1286, project Z01). This work was partially performed in Polifab, the micro and nanofabrication facility of Politecnico di Milano. 

\section*{Author contributions}
E.C. and D.K. conceived the idea and designed the experiments together with D.I. and C.T.. S.R. fabricated and characterized the volatile devices, carried out the \gls{wm} store and recall experiment, and analyzed the experimental data. D.K. ran the simulations of the large-scale simulation of \gls{wm} and of the associative symbolic \gls{wm}. The initial draft of the manuscript was written by E.C. and D.K.. All the authors discussed the results and provided feedback. E.C., D.I., and C.T. supervised the research.

\appendix

\renewcommand{\thesubsection}{S\arabic{subsection}}
\renewcommand{\theequation}{S\arabic{equation}}
\renewcommand{\thefigure}{S\arabic{figure}}
\setcounter{figure}{0}
\counterwithin*{equation}{section}

\section*{Supplementary Information}

\textbf{\huge{Tunable Synaptic Working Memory with Volatile Memristive Devices}}\\ \\

\author{Saverio Ricci*$^{1}$, David Kappel*$^{2,3}$, Christian Tetzlaff$^{2,4}$, Daniele Ielmini$^{1}$, and Erika Covi$^{5}$\\
\\
$^{1}$ Dipartimento di Elettronica, Informazione e Bioingegneria (DEIB), Politecnico di Milano, Piazza Leonardo da Vinci 32, 20133 Milano, Italy \\
$^{2}$ Bernstein Center for Computational Neuroscience,  III Physikalisches Institut – Biophysik, Georg-August Universität, Friedrich-Hund-Platz 1, 37077 Göttingen, Germany \\
$^{3}$ Institut für Neuroinformatik, Ruhr-Universität Bochum, Universitätsstr. 150 NB 3/32,
44801 Bochum, Germany \\
$^{4}$ Group of Computational Synaptic Physiology, Department for Neuro- and Sensory Physiology, University Medical Center G\"ottingen, Von-Siebold-Str. 3A, 37075 G\"ottingen, Germany \\
$^{5}$ NaMLab gGmbH, N\"othnitzer Str. 64 a, 01187 Dresden, Germany \\
* These Authors contributed equally to this work.}

\begin{figure}[ht]
    \centering
    \includegraphics[width=0.95\textwidth]{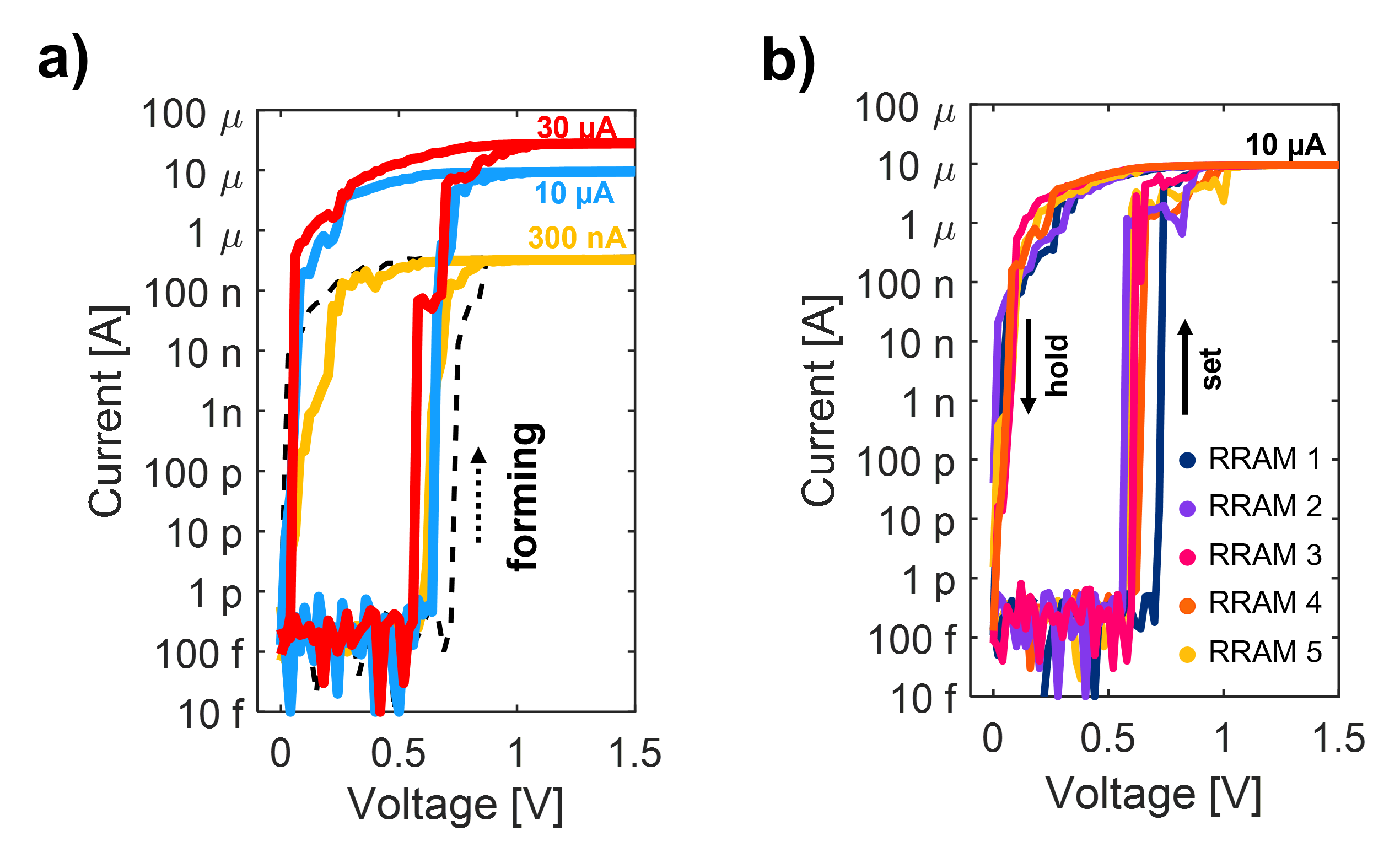}
    \caption{Quasi-static characterization of the devices. a) First cycle (dashed lines) and subsequent cycles at different compliance currents.  b) Comparison of 5 different devices.} 
    \label{fig:Supfig_01}
\end{figure}

Fig.~\ref{fig:Supfig_01}a reports an analysis of quasi-static measurements conducted on an Ag-based volatile memory. 
The characterization highlights that the forming (dashed line), or soft breakdown of the dielectric layer, has a voltage equal to the set ones and the set voltages are independent on the compliance currents. 

The conductance of the device, measured at 0.1\,V, passes from a \gls{hrs} $\ge$ 10\,T$\Omega$ to a \gls{lrs} $\approx$\,10\,k$\Omega$, with linear ohmic behavior in this last condition. The R$\mathrm{_{ON}}$ of the MOS transistor is in the order of a few hundreds ohm, well below the device resistance.

Fig.~\ref{fig:Supfig_01}b compares the I-V curves of the 5 devices used for the experiments, which show small device-to-device variability.

\newpage

\begin{figure}[ht]
    \centering
    \includegraphics[width=0.95\textwidth]{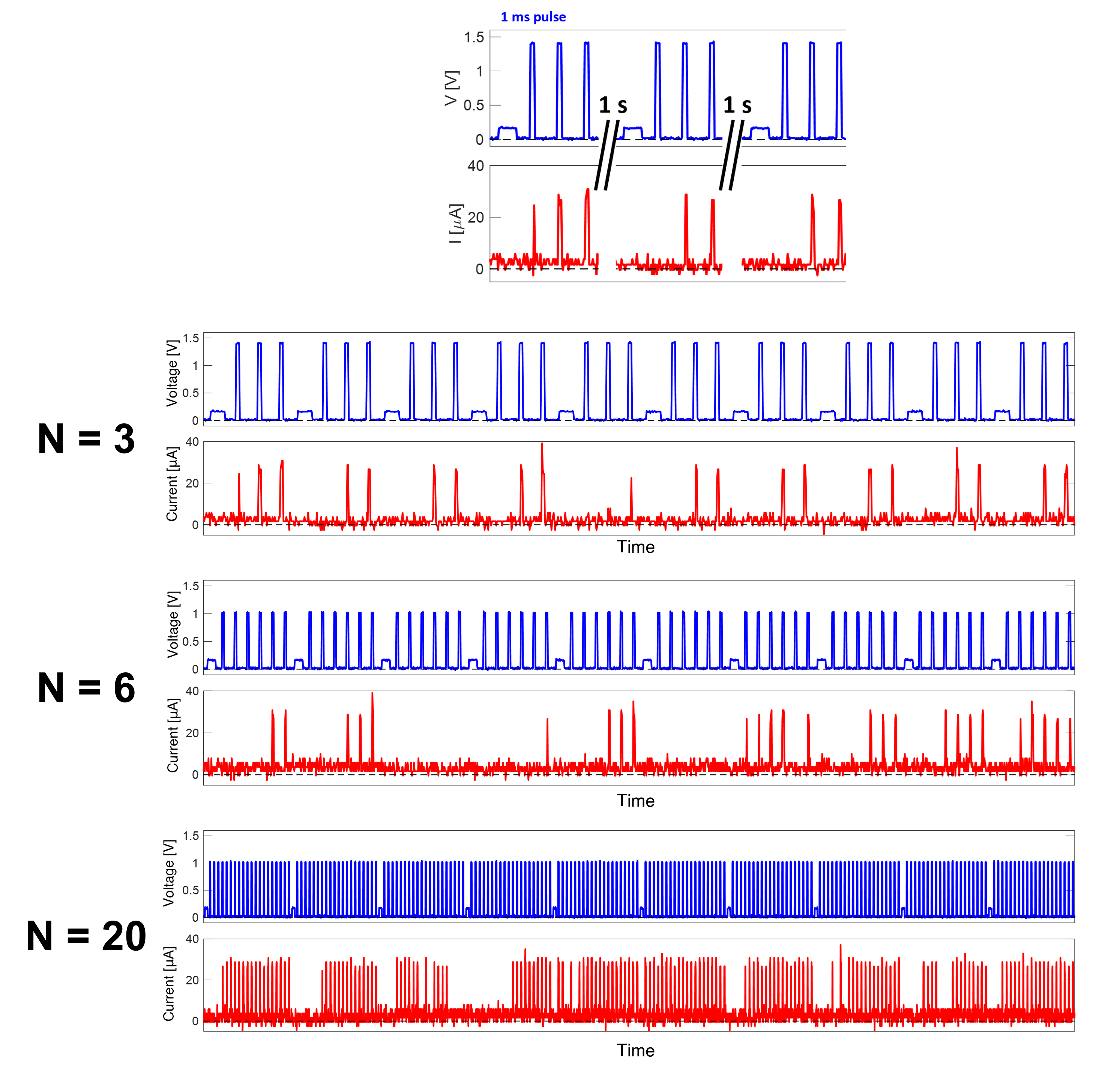}
    \caption{
    Switching of volatile memristive devices when stimulated with different number of pulses and voltage amplitudes. The trains of pulses are applied at 1\,s time interval. The presented images are cropped (as shown on top).}
    \label{fig:Supfig_02}
\end{figure}

To study the switching probability for a given voltage as a function of the number of pulses (P$_{ON}$(N)), trains of N 1\,ms pulses with a period of 11\,ms are applied. Each train is preceded by a 10\,ms reading pulse to verify that the device is initially in its \gls{hrs}. During this characterization, the I$_C$ was 17\,\textmu A, resulting in an average retention time of 28\,ms. Therefore, the trains of pulses are applied at 1\,s time intervals to allow the device to switch off. Fig.~\ref{fig:Supfig_02} compares the impact of the number of pulses (N) per train.

The smaller the voltage amplitude, the smaller the switching probability (P$_{ON}$(V)). For N = 6, for example, the device switches ON after the 4$^{th}$ pulse of the first train, whereas during the 3$^{rd}$ train the device does not switch ON. Increasing the number of pulses in the train to N = 20 increases the probability of the device to be in \gls{lrs} at the end of the train.

\newpage

\begin{figure}[ht]
    \centering
    \includegraphics[width=0.95\textwidth]{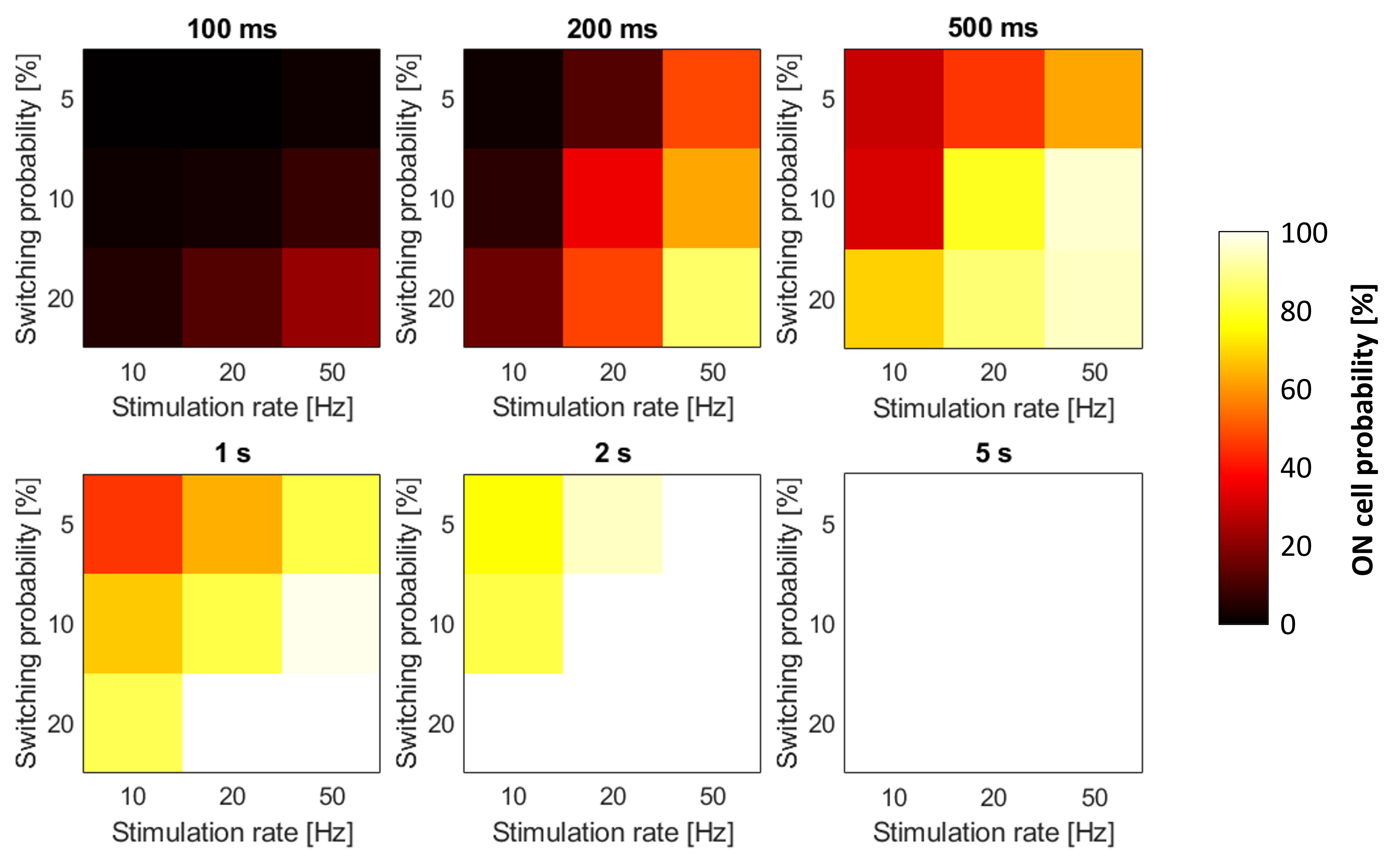}
    \caption{Probability of switching a device ON in a certain time window as a function of the stimulation rate and the switching probability.}
    \label{fig:Supfig_03bis}
\end{figure}

Fig.~\ref{fig:Supfig_03bis} compares different time windows in which a volatile device is continuously stimulated with a constant stimulation rate for a given switching probability.

Within the same time window, the probability of switching a device ON increases increasing either the switching probability (i.e., the voltage amplitude) or the stimulation rate, which is linked to the number of pulses in the train. As an example, with a time window of 200\,ms, the probability to find a device ON after stimulation ranges from almost 0\% with a stimulation rate of 10\,Hz and switching probability of 5\% to about 80\% with a stimulation rate of 50\,Hz and switching probability of 20\%.

If the stimulation rate and switching probability are fixed, longer time windows increase the probability of the device to be ON at the end of the stimulation sequence. If we consider a 10\,Hz stimulation with 5\% switching probability, the probability to find a device ON after stimulation ranges from almost 0\% with a time window of 100\,ms to 100\% with a time window of 5\,s. 

\newpage

\begin{figure}[ht]
    \centering
    \includegraphics[width=0.95\textwidth]{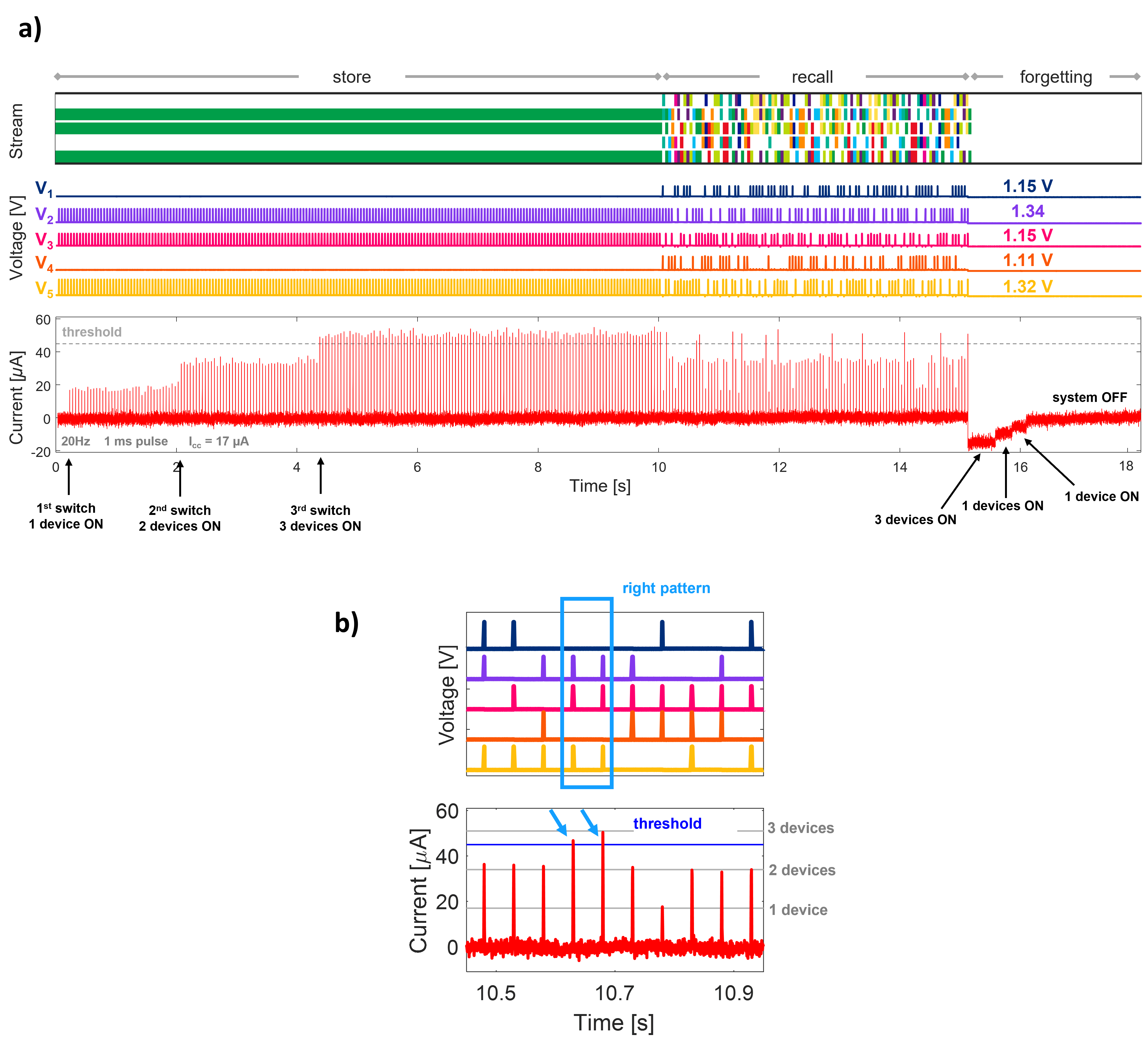}
    \caption{\gls{wm} store and recall experiment. a) The top panel shows the color input stream. Each color is encoded in a specific pattern (middle panel)
    The voltage amplitudes are chosen to have a switching probability of 5\%. The bottom panel shows the current collected from the devices that flows into the neuron. During the store phase the 3 devices that encode the color green switch on at different times. During the recall phase, the color is recognized if the current from the memristive synapses exceeds a threshold of 43\,\textmu A. In the forgetting phase, the 3 devices switches off at different times, thus indicating that the network forgot the stored element. b) Zoom of a part of the recall phase. Only the stored pattern, in the blue box, provides a current above the threshold, thus triggering the firing of the neuron.} 
    \label{fig:Supfig_03}
\end{figure}

Fig.~\ref{fig:Supfig_03} reports the example of an experimental trace, presenting the store, recall, and forget phases. 
During the store phase, a single pattern (the green one in the example, encoded as 01101) is applied multiple times and until the three devices associated with 1 are in \gls{lrs}. The voltages applied to the devices and shown in the middle panel of Fig.~\ref{fig:Supfig_03}a are chosen to have a 5\% switching probability (P$_{ON}$). As a consequence, the total current of the system increases (bottom panel). Since the devices have all the same current compliance of 17\,\textmu A, each device that switches on provides a contribution in current of 17\,\textmu A.

During the recall phase, random patterns are fed to the system. The current contribution of each stimulated device in \gls{lrs} is 17\,\textmu A, therefore we expect 51\,\textmu A when 3 ON devices are stimulated (stored pattern) and 34\,\textmu A when 2 ON devices are stimulated (worst case for the non-stored patterns). Therefore, a threshold of 42\,\textmu A is set. When the stored pattern is presented to the network, the current contribution exceeds the threshold, thus triggering the fire of the neuron. The current contribution of the other patterns is lower than the threshold, as better highlighted in Fig.~\ref{fig:Supfig_03}b.

After the recall phase, the system is monitored in absence of stimulation with a reading voltage of -150\,mV applied to all the devices. After about 1\,s, all the devices switch off, thus indicating that the network forgot the stored element. As in the store phase the devices have different switching on times, also the switch off occurs with different times due to the stochastic nature of the \gls{cf}.

\newpage

\begin{figure}[ht]
    \centering
    \includegraphics[width=0.8\textwidth]{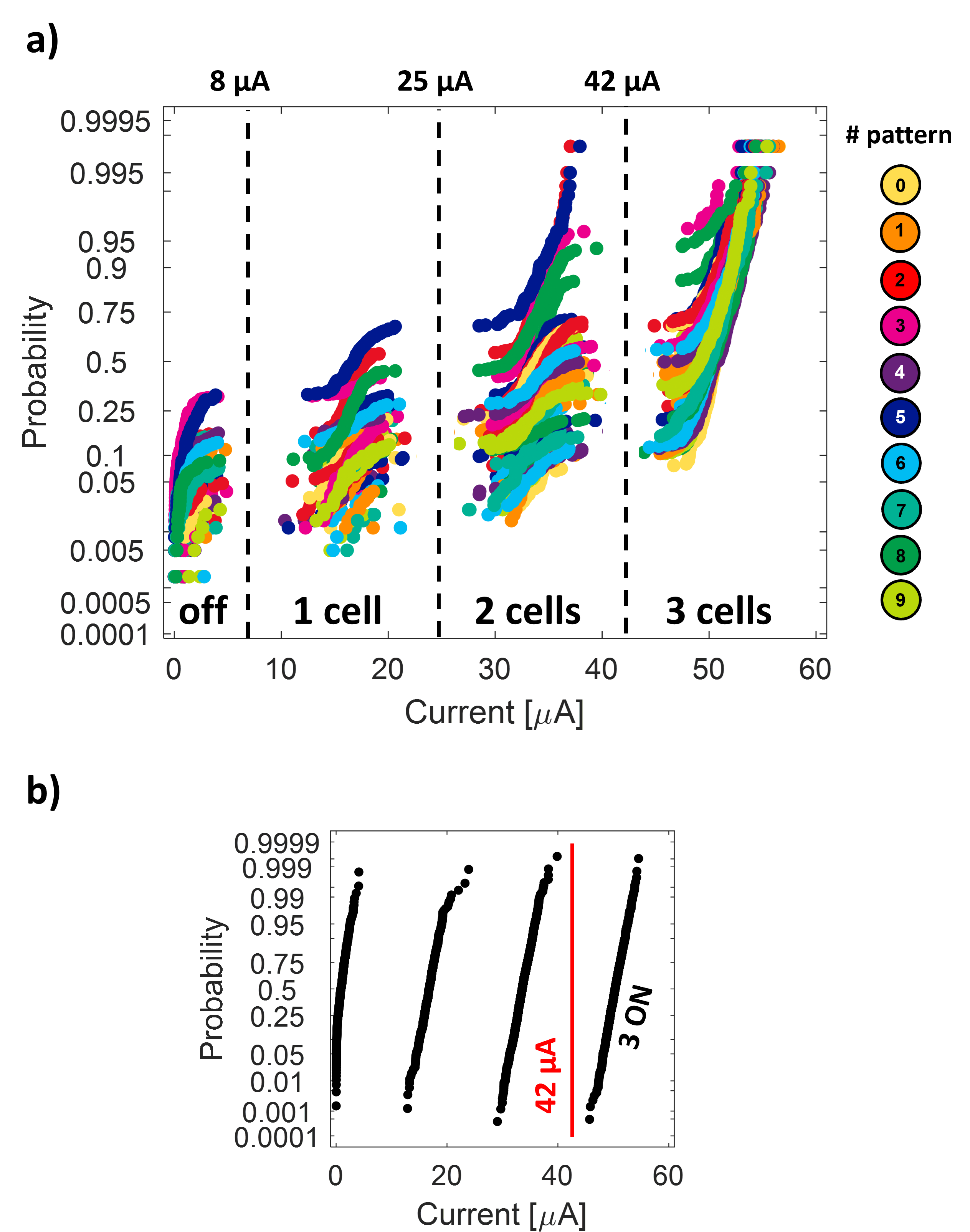}
    \caption{Currents distribution of the 5 devices used in the store and recall experiment. a) Current read when all devices are in \gls{hrs} and when one, two, or three are in \gls{lrs}. Data grouped by stored pattern. The dashed lines separate number of ON devices. 
    b) Current distribution of the currents from all the patterns combined. 42\,\textmu A threshold is set as threshold.}
    \label{fig:Supfig_04}
\end{figure}

A threshold current is used to recognize the stored pattern from the other. According to the current levels presented in Fig.~\ref{fig:Supfig_04}, a 42\,\textmu A threshold is suitable to distinguish the stored pattern, since there is a clear non-overlap among the different current distributions.

\newpage

\begin{figure}[ht]
    \centering
    \includegraphics[width=0.95\textwidth]{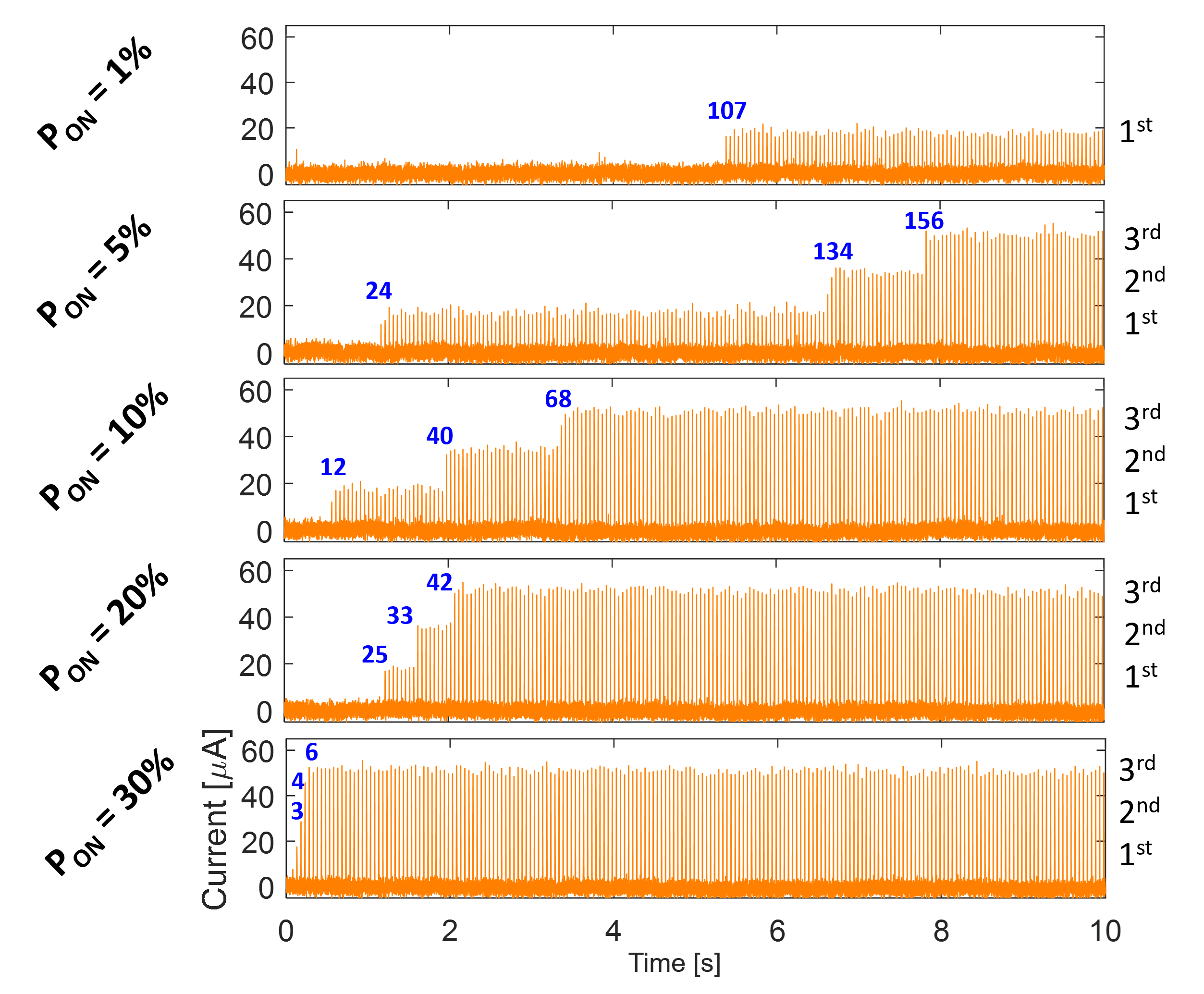}
    \caption{Impact of the switching probability on the time needed to store a pattern. The stimulation rate is with 20\,Hz, whereas the switching probability is changed from 1\% to 30\%. The blue numbers inside each plot indicate how many pulses are necessary before a device is set to \gls{lrs}.}
    \label{fig:Supfig_05}
\end{figure}
 
Fig.~\ref{fig:Supfig_05} shows the effect of the switching probability on the store phase. Low switching probabilities require long trains of pulses to set the stimulated devices in \gls{lrs}. As an example, with P$\mathrm{_{ON}}$\,=\,1\%, 10\,s stimulation are not enough to store a pattern and only one devices is in \gls{lrs} by the end of the stimulation. An increase of P$\mathrm{_{ON}}$ shortens the time needed to store a pattern, which can be don to less than 1\,s in the case of P$\mathrm{_{ON}}$\,=\,30\%. It should be noted, however, that a high P$\mathrm{_{ON}}$ might be beneficial in storing a pattern in a short time, but in the recall phase leads to an early forgetting of the pattern. It is therefore important to select the P$\mathrm{_{ON}}$ while considering all the phases of the \gls{wm} system.

\newpage

\begin{figure}[ht]
    \centering
    \includegraphics[width=0.95\textwidth]{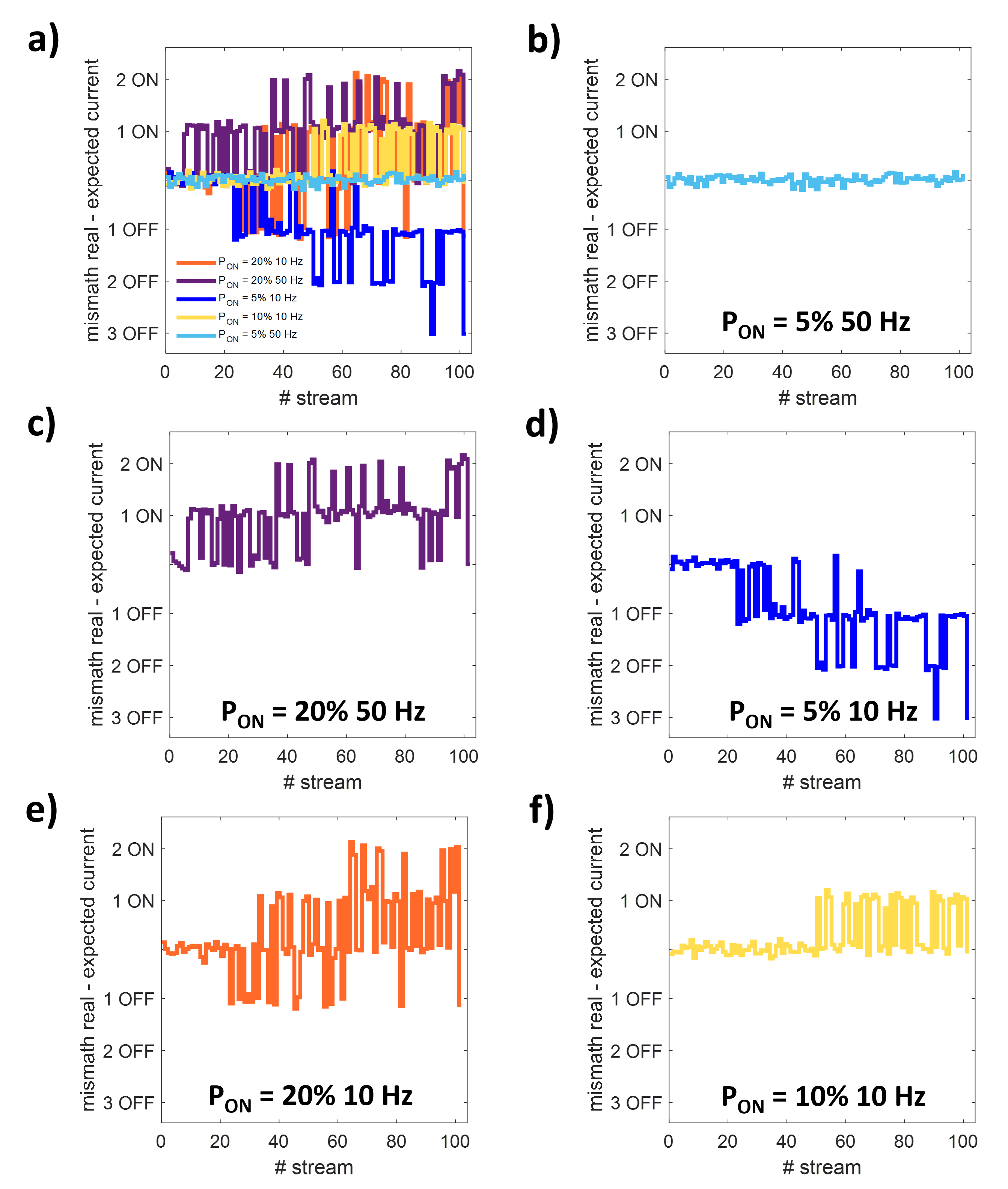}
    \caption{Failure analysis during the recall phase of the \gls{wm} store and recall experiment. a) Example of failure defined as the mismatch between the number of actual ON devices and the number of expected ones when a pattern is presented to the network. b-f) Single plots of the data in a) to highlight the behavior of the \gls{wm} system under different P$\mathrm{_{ON}}$ and stimulation rates.}
    \label{fig:Supfig_06}
\end{figure}

The possible failures of the system can be measured by comparing the expected and read currents, which correspond to the expected and actual number of devices in \gls{lrs} contributing to the read current (Fig.~\ref{fig:Supfig_06}a). Four possible situations can arise, summarized in Fig.~\ref{fig:Supfig_06}:
\begin{itemize}
    \item No difference between expected and read current (Fig.~\ref{fig:Supfig_06}b), no failures.
    \item Expected current lower than read current (Fig.~\ref{fig:Supfig_06}c and f). This is usually the case when the switching probability and/or the stimulation rate are too high. One or more devices, previously in \gls{hrs}, are switched to \gls{lrs} during the recall phase.
    \item Expected current greater than read current (Fig.~\ref{fig:Supfig_06}d). In this case case the switching probability and/or the stimulation rate are too low and one or more devices, previously in \gls{lrs}, switch back to \gls{hrs} during the recall phase.
    \item Combination of the two previous cases (Fig.~\ref{fig:Supfig_06}e). This is an unstable situation in which during the recall phase some devices switch from \gls{lrs} to \gls{hrs} and vice versa. In this case, the stochastic nature of the memristive devices is the main reason of the fluctuations observed.
\end{itemize}

\newpage

\begin{figure}[ht]
    \centering
    \includegraphics[width=0.95\textwidth]{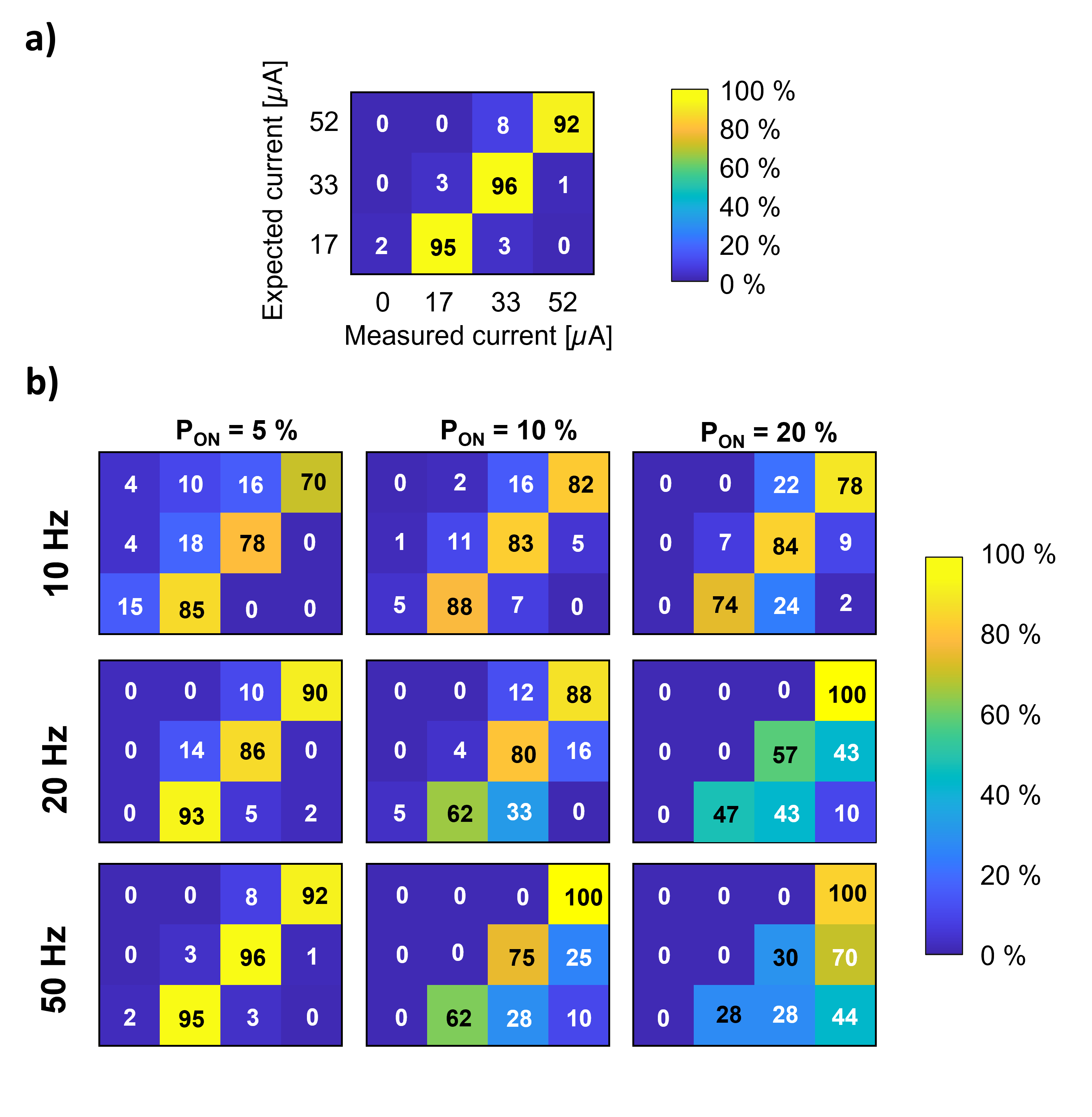}
    \caption{Correlation plots of the system response. a) The measured current is shown on the horizontal axis and the expected current is on the vertical axis. The color-bar, as well as the numbers inside each square indicate the probability of each occurrence. b) Correlation plots for different experimental conditions.}
    \label{fig:Supfig_07}
\end{figure}

Fig.~\ref{fig:Supfig_07} shows the correlation plot of different experiments carried out by varying the switching probability and the repetition rate. Lower P$\mathrm{_{ON}}$ improves the classification accuracy of the system. When the stimulation rate is lower than the average retention time of the device, the devices tend to switch from \gls{lrs} to \gls{hrs} during the recall phase, thus leading to higher probability of forgetting the pattern. High stimulation rates instead, cause devices to switch from \gls{hrs} to \gls{lrs}, thus overwriting the stored pattern. A careful choice of average retention times and stimulation rate has therefore to be made to maximize the system accuracy.

\newpage

\begin{figure}[ht]
    \centering
    \includegraphics[width=0.95\textwidth]{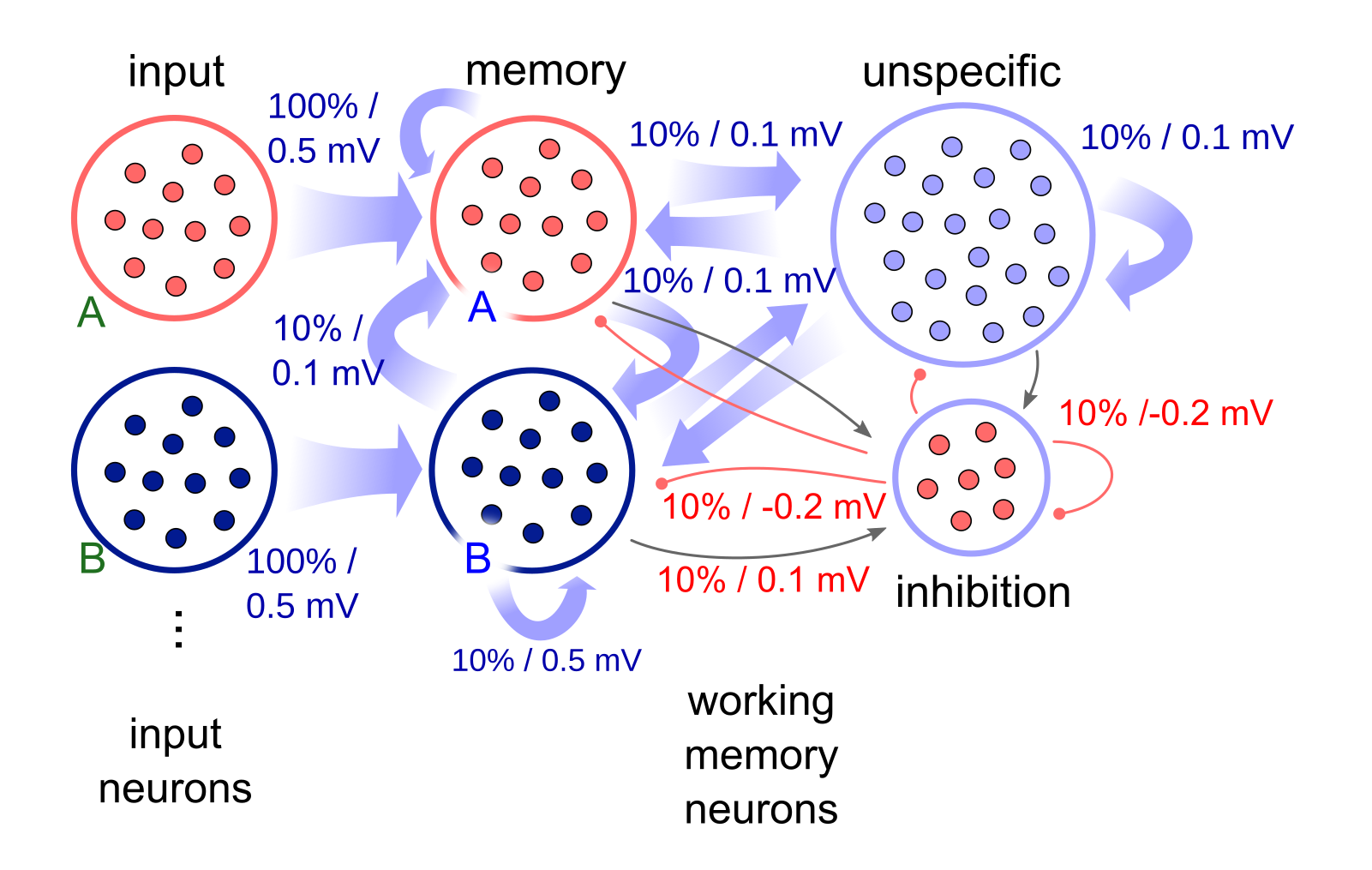}
    \caption{Details to the working memory network model. Connection probabilities (in \%) and PSP amplitude shown between neuronal populations. Input and working memory populations only shown for two memory items to keep the figure uncluttered.}
    \label{fig:Supfig_08}
\end{figure}

\newpage

\begin{figure}[ht]
    \centering
    \includegraphics[width=0.95\textwidth]{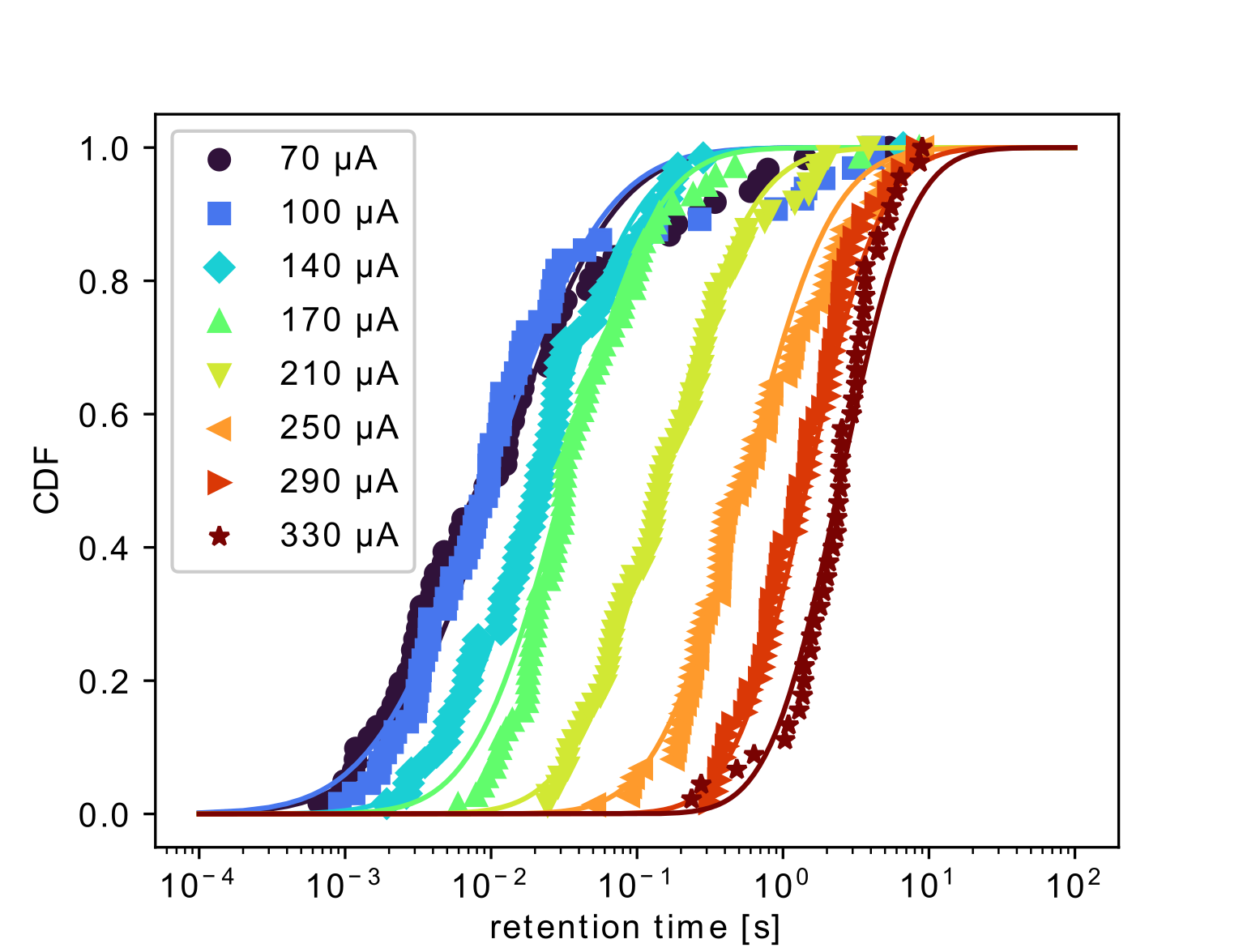}
    \caption{Lognormal fits (solid lines) to empirical cumulative distribution of retention times for different currents (data points).}
    \label{fig:Supfig_lognormal_fit}
\end{figure}

Fig.~\ref{fig:Supfig_lognormal_fit} shows the lognormal fit to the experimental data of the memristive device. The device is stimulated with semi-triangular pulses as in Fig.~\ref{fig:fig2} and the current compliance was varied by changing the voltage at the gate of the transistor.

\end{document}